\documentclass[prd,aps,showpacs,tightenlines, 10pt, reprint,nofootinbib,superscriptaddress,floatfix,twocolumn]{revtex4-2}

\usepackage[utf8]{inputenc} 

\usepackage{acronym}
\usepackage{amsmath}
\usepackage{amsthm}
\usepackage{amssymb}
\usepackage{mathtools} 
\usepackage{phaistos}
\usepackage{graphicx}
\usepackage{wrapfig}
\usepackage{subcaption}
\usepackage[section]{placeins}

\usepackage{physics}
\usepackage{IEEEtrantools}
\usepackage{xspace}
\usepackage{cancel}

\usepackage{epsfig}
\usepackage{multirow} 
\usepackage{color}
\usepackage[dvipsnames]{xcolor}
\usepackage{rotating}
\usepackage[export]{adjustbox}
\usepackage{floatrow}

\newfloatcommand{capbtabbox}{table}[][\FBwidth]
\usepackage{tabularx}
\usepackage[pdfencoding=auto, psdextra]{hyperref}
\usepackage{units}
\usepackage[T1]{fontenc}
\usepackage[normalem]{ulem}
\usepackage{orcidlink}

\DeclarePairedDelimiterX{\avg}[1]{\langle}{\rangle}{#1}
\makeatletter
\let\oldavg\avg
\def\avg{\@ifstar{\oldavg}{\oldavg*}}  

\renewcommand{\eqref}[1]{Eq.~(\ref{#1})}

\begin{document}

\title{Forecasted Detection Limits on the (Dark) Matter Density in Supermassive Black Hole Binaries for LISA}
\author{Matthias Daniel\,\orcidlink{0009-0001-5805-2802}}
\email{daniel@itp.uni-frankfurt.de} 
\affiliation{Institute for Theoretical Physics, Goethe University, 60438 Frankfurt am Main, Germany}

\author{Kris Pardo\,\orcidlink{0000-0002-9910-6782}}
\email{kmpardo@usc.edu}
\affiliation{Department of Physics and Astronomy, University of Southern California, Los Angeles, CA 90089, USA}

\author{Laura Sagunski\,\orcidlink{0000-0002-3506-3306}}
\email{sagunski@itp.uni-frankfurt.de} 
\affiliation{Institute for Theoretical Physics, Goethe University, 60438 Frankfurt am Main, Germany}

\begin{abstract}
\noindent Supermassive black hole binaries (SMBHBs) are among the most powerful known sources of gravitational waves (GWs). Accordingly, these systems could dominate GW emission in the micro- and millihertz frequency range. Within this domain, SMBHs evolve rapidly and merge with each other. Dynamical friction from stars and gas at the centers of galaxies typically helps to bring together two SMBHs when they are at relatively far separations ($\approx \rm{kpc} - 100~\rm{pc}$), but becomes less efficient at smaller separations. However, dark matter (DM) spikes around SMBHs could enhance dynamical friction at close separations and, thus, shorten the evolution times. In this paper, we simulate the effects of DM spikes on GW signals in the micro- to millihertz frequency range and confirm that the GW signals from SMBHBs with DM spikes can be clearly distinguished from those without any additional matter. Making use of the projected sensitivity curve of the Laser Interferometer Space Antenna (LISA), we forecast upper limits for the (dark) matter density for given future SMBHB observations. We then compare these thresholds with the theoretical density profiles expected for self-interacting dark matter (SIDM) spikes.
\\[0.3cm]
\textit{Keywords: supermassive black hole binaries - gravitational waves - dark matter - LISA}
\end{abstract}
\preprint{}
\maketitle

\section{Introduction \label{sec:intro}} 
The existence of gravitational waves (GWs) was experimentally confirmed in 2015 by the direct observation of GWs from a binary black hole merger, GW150914, with the two LIGO detectors \cite{Abbott_2016_2}. Since then, LIGO, along with the Virgo and KAGRA detectors, has collected a sizable amount of data from mergers of binary neutron stars and stellar-mass black holes \cite{LIGO+VIRGOCollaboration_2023}. As GW detectors based on the principle of a Michelson interferometer, it is difficult to protect these ground-based detectors from seismic noise and the fluctuating gravitational potentials caused by environmental factors on Earth for frequencies below approximately $10$ Hz \cite{Abbate_2003}. Therefore, to detect low-frequency GW signals, especially within the $10^{-5}-10^0$ Hz range, future space-based detectors, such as LISA, are necessary \cite{Amaro-Seoane_2017}. The GWs produced by the inspiral and coalescence of supermassive black holes (SMBHs) in binary systems are expected to be the main signal in these frequencies and below. These binary systems are assumed to be formed through the merger of two massive galaxies, each harboring a SMBH with a mass of $10^5 - 10^{10}\, \mathrm{M_{\odot}}$ in its central region \cite{Maggiore_V2_2008, Celoria_2018}.

Given the frequent observation of galaxy mergers \cite{Bell_2006} and the enormous number of galaxies in the observable universe, it is assumed that a substantial cosmic population of supermassive black hole binaries (SMBHBs) contribute to the production of a stochastic gravitational wave background (GWB) across a wide frequency range. In their 15-year data set, the NANOGrav collaboration found strong evidence for the presence of a stochastic GWB in the nanohertz frequency range, which could be generated by the incoherent superposition of GWs from a cosmic population of inspiraling SMBHBs \cite{Agazie_2023_1, Agazie_2023}. This has recently been substantiated by other Pulsar Timing Array (PTA) collaborations, including EPTA, PPTA and CPTA \cite{Antoniadis_2023,PPTA_2023,Xu_2023}.

Despite these strong indications of SMBHB existence \cite{Alachkar_2023, Rodriguez_2009}, their formation, evolution, interaction with their host galaxies, and the nature and properties of their matter environment are not yet understood. To clarify these fundamental questions, astrophysical information about the surrounding matter can be extracted from the GW signals emitted by these systems \cite{Ghoshal_2023,Shen_2023,Bonetti_2023, Ellis_2023, Bi_2023}. There are many different structures made of matter that can form around black holes (BHs) and influence their GW emission. In the following, a brief overview of the theoretical results on that is provided for some of these structures.

Refs.~\cite{Eda_2013, Eda_2015} were the first to investigate the influence of dark matter (DM) spikes on intermediate mass ratio inspirals\footnote{IMRIs are the inspirals of a compact object with stellar mass (black hole or neutron star) toward an intermediate mass black hole with a mass ranging from $10^3$ to $10^5\,\mathrm{M_\odot}$.} (IMRIs) and their GW signals. They found that the detectability of the effect of DM spikes on the GW spectrum  by LISA strongly depends on the radial density profile of the spikes. DM spikes with slopes greater than $\alpha \simeq  1.7$ have a significant impact on the observed gravitational waveform and can therefore be seen in the GW signal. Subsequent work has explored circular and eccentric IMRIs with static DM spikes considering different spike slopes \cite{Yue_2019,Becker_2022}, as well as binaries that counter-rotate or co-rotate with respect to the DM spike particles \cite{Mukherjee_2023}.

In addition, there are also studies that investigated the influence of accretion disks compared to the effect of DM spikes on the dynamical evolution of binary systems and their GW signals. Both DM spikes and accretion disks around BHs result in the loss of orbital energy and angular momentum, e.g., through dynamical friction \cite{Chandrasekhar_1943}, leading to a gradual approach of the two BHs. To be more precise, DM spikes, driving IMRIs, leave different fingerprints in the dephasing and breaking index of their GW signals compared to accretion disks composed of baryonic or dark matter \cite{Yue_2018, Becker_2023, Nichols_2023}. By performing a Bayesian analysis, Ref.~\cite{Cole_2022} showed that for IMRIs, these different environmental effects can even be easily distinguished from each other with LISA.

Furthermore, in numerous studies, DM is also considered as a possible solution of the final parsec problem \cite{Milosavljevic_2003} in binary systems of different mass scales. On the one hand, the existence of halos or solitons composed of wave-like, ultra-light DM around black holes could provide a plausible explanation for this problem \cite{Koo_2023, Bromley_2023}. On the other hand, the final parsec problem in SMBHBs is solved by dynamical friction with SIDM particles possessing cross sections per unit mass on the order $\mathrm{cm^2/g}$ \cite{Alonso-Alvarez_2024, Fischer:2024dte}. In both scenarios, the presence of DM also leaves an imprint on the characteristic strain of the stochastic GWB in the nanohertz frequency range \cite{Aghaie_2024}. Because the environment of binary systems can affect the GW signal, we can use GWs to constrain the environmental composition and dynamic properties of binary systems.

Ref.~\cite{Coogan_2022} studied the detectability of IMRIs occurring within isotropic DM spikes evolving over time. Their findings showed that such IMRIs with chirp masses greater than $\sim 16\,\mathrm{M_{\odot}}$ can be detected by LISA up to a luminosity distance of $\sim 75\,\mathrm{Mpc}$ with a signal-to-noise ratio of $\gtrsim 15$. To get these results, the authors considered the same parametrization for the initial DM spike profiles as we do to describe our time-independent spikes (see \ref{subsec:dm-spike}). In contrast, this work focuses on the detectability of individual SMBHBs surrounded by any static and spherically symmetric matter density profile. We forecast upper bounds for the (dark) matter density in the environment of SMBHs across LISA's frequency range for different black hole masses and redshifts. We then compare our results with the theoretical density distributions expected for SIDM spikes \cite{Gondolo_1999, Tulin_2018}.

We structure this paper as follows. In Sec.~\ref{sec:theory}, we describe the theoretical framework for modeling the orbital evolution of SMBHBs and their GW signals. In Sec.~\ref{sec:results}, we present and discuss the results. We conclude in Sec.~\ref{sec:conclusions}.

Our numerical code is publicly available under \url{https://github.com/DMGW-Goethe/SMBHBpy} (Ref.~\cite{Daniel_SMBHBpy}).

\section{SMBHB Modeling \label{sec:theory}}
In this section, we provide a description of the orbital evolution of SMBHBs under the influence of GW emission and dynamical friction with DM particles. We then discuss the form of the GW signals emitted by individual binaries. Finally, we provide the mathematical relationship between the matter density around a SMBHB and the characteristic strain of the corresponding GW signal.

The constituents of the SMBHBs we consider lose energy and angular momentum due to the emission of GWs and via the dynamical friction mechanism through their interaction with DM. For the results in \ref{subsec:results_1}, we model the SMBHs as Schwarzschild BHs, each surrounded by a spherically symmetric and static DM spike. Within the framework of the Newtonian approximation, the SMBHs are treated as point-like particles with masses $m_1$ and $m_2$, moving on eccentric or circular Keplerian orbits around their common center of mass. These elliptical orbits can be described by the semi-major axis $a$, the eccentricity $e$ and the eccentric anomaly $\phi$. The mass ratio and spike parameters of both SMBHs can be chosen arbitrarily. If the theoretical corrections to the Newtonian description of the trajectories of both SMBHs due to the effects of general relativity become too significant, i.e., for distances between the SMBHs that are smaller than the sum of their innermost stable circular orbits (ISCOs), the numerical calculation of the SMBHB evolution in our code is terminated. Thus, we can employ a Newtonian description of the SMBHB evolution here.

\vspace{-0.25cm}
\subsection{DM spike density profile \label{subsec:dm-spike}}
When a BH undergoes adiabatic growth within a cuspy DM halo, the DM density within the sphere of influence of the BH with spatial extent $r_h$ can significantly increase \cite{Gondolo_1999, Lacroix_2018}. The density profile of the resulting DM spike can be described by a simple power-law:
\begin{align}
    \rho_{\text{DM}}(r) = \begin{cases}
                        0 & \text{, } r < r_{\text{min}} \\
                        \rho_{\text{sp}} \left(\frac{r_{\text{sp}}}{r}\right)^{\alpha} & \text{, } r_h \geq r \geq r_{\text{min}} \\
                      \end{cases},
    \label{eq:rho_DM(r)}
\end{align}
\noindent where $\rho_{\text{sp}}$ is the normalization density, $r$ is the distance from the BH, $\alpha$ denotes the slope of the profile and $r_{\text{sp}}$ is the parameter for the spike size, which can be empirically defined by $r_{\text{sp}} \approx 0.2\, r_h$ \cite{Eda_2015} with
\begin{align}
    r_h = \frac{G\, m_{\text{BH}}}{\sigma^2} \; .
    \label{eq:r_h}
\end{align}
\noindent Here, $G$ denotes Newton's gravitational constant, $m_{\text{BH}}$ is the mass of the Schwarzschild BH, and $\sigma$ represents the velocity dispersion of the host galaxy-bulge.

In this work, we equate $r_{\text{min}}$ in Eq.~\ref{eq:rho_DM(r)} with the ISCO of a Schwarzschild BH, i.e., 
\begin{align}
    r_{\text{min}} = r_{\text{ISCO}} = 6\, \frac{G\, m_{\text{BH}}}{c^2} \; .
    \label{eq:r_min}
\end{align}
\noindent where $c$ is the speed of light in vacuum.

The slope $\alpha$ of the spike profile can take different values depending on the slope $\alpha_{\text{ini}}$ of the initial halo according to $\alpha = (9-2 \alpha_{\text{ini}})/(4-\alpha_{\text{ini}})$ \cite{Gondolo_1999}. For example, an initial Navarro-Frenk-White (NFW) halo consisting of cold dark matter (CDM) particles with $\rho_{\text{NFW}}(r) \propto r^{-1}$ for small $r$ leads to a corresponding slope for a CDM spike of $\alpha_{\text{CDM}} = 7/3$ \cite{Navarro_1996}. For a SIDM spike, on the other hand, $\alpha_{\text{SIDM}} = (3+b)/4$, with the parameter $b = 0, 1, \dots, 4$ characterizing the type of the particle interactions \cite{Shapiro_2014, Alonso-Alvarez_2024}. For instance, $b = 0$ describes isotropic scattering, arising from contact interactions, whereas $b=4$ is associated with the Coulomb scattering, mediated by a massless force carrier \cite{Alonso-Alvarez_2024}.

At this point, it should be noted that the formation and evolution of DM spikes around SMBHs in binary systems are still not fully understood. However, Refs.~\cite{Kavanagh_2020, Mukherjee_2023, Kavanagh_2025} suggest that their temporal stability strongly depend on the specific properties of both the binary, such as the mass ratio, and the spike itself. We will discuss this in more detail in \ref{subsec:results_1} and present our corresponding computation in the Appendix \ref{sec:energy_balance_CDM}.

The total mass of DM within the sphere of influence around a BH is approximately given by \cite{Eda_2015} 
\begin{align}
    M_{\text{DM}}(r < r_h) \approx 4 \pi \int_{0}^{r_h} \rho_{\text{sp}} \left(\frac{r_{\text{sp}}}{r}\right)^{\alpha} \, r^2 \, dr \approx 2 \,m_{\text{BH}} \; .
    \label{eq:M_DM}
\end{align}
\noindent From this definition, we find
\begin{align}
    \rho_{\text{sp}} \approx \frac{(3-\alpha)\, 0.2^{3-\alpha}\, m_{\text{BH}}}{2 \pi {r_{\text{sp}}}^3}\; .
    \label{eq:rho_sp}
\end{align}

\begin{figure}
    \centering
    \includegraphics[width=0.98\columnwidth]{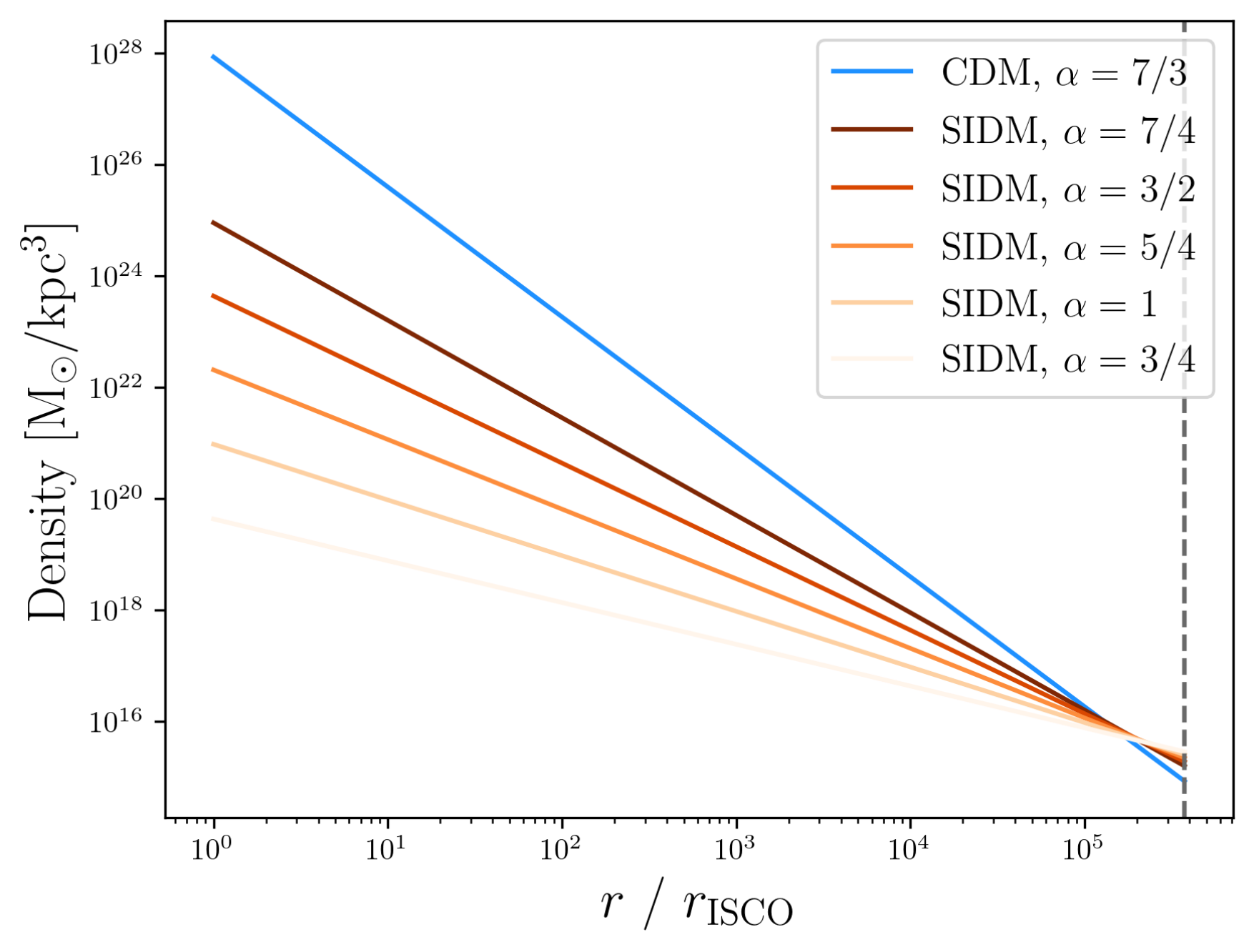}
    \caption{Radial density profiles for CDM ($\alpha = 7/3$) and SIDM spikes with different slopes, assuming $m_\mathrm{BH} = 10^7\,\mathrm{M_\odot}$ and $\sigma = 200\,\mathrm{km/s}$. The vertical dashed gray line indicates the influence radius $r_h$.}
    \label{fig:rho_vs_r_CDM+SIDM}
\end{figure}

Fig.~\ref{fig:rho_vs_r_CDM+SIDM} shows the corresponding DM density profiles for all the aforementioned slopes. It can be seen that lower values of $\alpha$ result in lower densities in the inner spike region.

In addition, it should be noted that mergers with smaller BHs or adiabatic growth of one of the SMBHs, which may not occur precisely at the center of its initial DM halo, can result in the formation of a weaker cusp \cite{Nakano_1999, Merritt_2002}. However, we do not consider these dynamic processes or other changes in the DM density distribution around SMBHs here.

Furthermore, we point out that a fully relativistic treatment of a DM spike results in the spike extending closer to the SMBH, down to $4Gm_\mathrm{BH}/c^2$, accompanied by an increase in density in the central region \cite{Sadeghian_2013}. The precise implications of this for our results will be briefly discussed in Sec.~\ref{sec:conclusions}.

\subsection{Orbital evolution \label{subsec:orbitalevolution}}

\subsubsection{Dynamical friction with DM spikes \label{subsubsec:DF}}
We now consider the effect of dynamical friction from DM spikes on SMBHBs. In what follows, the interaction of DM particles within the overlap region of the spikes between both SMBHs and the spike's proper motion relative to the corresponding SMBH are neglected.

Note that the DM density of a SMBH's own spike is zero, according to Eq.~\ref{eq:rho_DM(r)}. Consequently, only the DM surrounding the SMBH with mass $m_j$ has an influence on the motion of the other SMBH with mass $m_i$  ($i,j = 1,2; \, i \neq j$). The dynamical friction force $F_{\text{DF},i}$ acting on the $i$-th SMBH due to the DM spike around the $j$-th SMBH is given by Chandrasekhar's formula \cite{Chandrasekhar_1943}
\begin{align}
    F_{\text{DF},i} = 4\, \pi \,G^2\, m_i^2\, \rho_{\text{DM},j}({r})\, \ln(\Lambda) \,\frac{1}{v^2} \; ,
    \label{eq:F_DF}
\end{align}
\noindent where $v$ is the relative velocity of the binary system. Since this work is primarily concerned with the calculation of general upper limits on the (dark) matter density in SMBHBs for LISA, an order of magnitude estimate of the Coulomb logarithm $\ln(\Lambda)$ is sufficient. Therefore, we fix $\ln(\Lambda) = 10$ in the following \cite{BinneyTremaine_2008, Amaro_Seoane_2023}.

The energy loss of a SMBHB due to dynamical friction can be obtained as follows:
\begin{equation}
\begin{aligned}
    \frac{dE_{\text{DF}}}{dt} &=-\underbrace{\left(F_{\text{DF},1} \, \frac{m_2}{M} + F_{\text{DF},2} \, \frac{m_1}{M} \right)}_{=:\, F_{\text{DF}}} v\\[0.3cm]
    &= - 4 \pi G^2 \ln(\Lambda)  \frac{\mu}{v} \left(m_1\, \rho_{\text{DM},2}({r}) + m_2\, \rho_{\text{DM},1}({r})\right) \; .
\end{aligned}
\label{eq:dE_DF_}
\end{equation}
\noindent Here, the reduced mass $\mu = m_1 m_2/M$ was defined, with $M = m_1 + m_2$ being the total mass of the binary system.

For a dissipative force, such as the dynamical friction force $F_{\mathrm{DF}}$, the temporal average over an orbital period $T$ can be transformed into an integral over $\phi$ \cite{Maggiore_V1_2007},
\begin{equation}
   \left< \left(...\right) \right> = \int_{0}^{T} \frac{dt}{T} \left(...\right) = \left(1-e^2\right)^{3/2} \int_{0}^{2 \pi} \frac{d\phi\,\left(...\right)}{2 \pi \left(1+e\,\cos(\phi)\right)^{2}} \; . 
    \label{eq:transition_t_phi}
\end{equation}
\noindent The resulting energy and angular momentum loss through the interaction with the DM spikes is

\begin{equation}
\begin{aligned}
    \left< \frac{dE_{\text{DF}}}{dt} \right> &= - \left(1-e^2\right)^{3/2} \int_{0}^{2 \pi} \frac{d\phi}{2 \pi} \frac{F_{\text{DF}}(r(\phi),v(\phi))  v(\phi)}{\left(1+e\cos(\phi)\right)^{2}}\;,
\end{aligned}
\label{eq:dE_DF}
\end{equation}
\begin{equation}
\begin{aligned}
    \left< \frac{dL_{\text{DF}}}{dt} \right> &= - \frac{\sqrt{G M a}}{ \left(1-e^2\right)^{-2}} \int_{0}^{2 \pi} \frac{d\phi}{2 \pi} \frac{F_{\text{DF}}(r(\phi),v(\phi))}{\left(1+e\cos(\phi)\right)^{2} v(\phi)}\;.
\end{aligned}
\label{eq:dL_DF}
\end{equation}

\subsubsection{GW emission \label{subsubsec:GW-emission}}
The energy and angular momentum loss of the binary system due to GW emission are given by \cite{Maggiore_V1_2007, Peters_1963, Peters_1964}

\begin{align}
    \left< \frac{dE_{\text{GW}}}{dt} \right> = - \frac{32}{5} \frac{G^4 \mu^2 M^3}{c^5 a^5} \frac{1+\frac{73}{24} e^2 + \frac{37}{96} e^4}{\left(1-e^2\right)^{7/2}} \; ,
    \label{eq:dE_GW}
\end{align}
\begin{align}
    \left< \frac{dL_{\text{GW}}}{dt} \right> = - \frac{32}{5} \frac{G^{7/2} \mu^2 M^{5/2}}{c^5 a^{7/2}} \frac{1+\frac{7}{8} e^2}{\left(1-e^2\right)^2} \; .
    \label{eq:dL_GW}
\end{align}

\subsubsection{Energy and angular momentum balance \label{subsubsec:E-L-balance}}
Finally, the energy and angular momentum balance of an individual SMBHB can be formulated as
\begin{align}
    \frac{dE_{\text{orb}}}{dt} = \left< \frac{dE_{\text{DF}}}{dt} \right> + \left< \frac{dE_{\text{GW}}}{dt} \right> \; ,
    \label{eq:dE_orb}
\end{align}
\begin{align}
    \frac{dL_{\text{orb}}}{dt} = \left< \frac{dL_{\text{DF}}}{dt} \right> + \left< \frac{dL_{\text{GW}}}{dt} \right> \; .
    \label{eq:dL_orb}
\end{align}

In the initial phase of SMBHB evolution, dynamical friction is the dominant source of energy loss. However, the final phase of the inspiral is dominated by GW emission. This is because the dynamical friction force due to the interaction with DM particles is proportional to $v^{-2}$ (compare Eq.~\ref{eq:F_DF}), so dynamical friction becomes negligible above a certain relative velocity between the two SMBHs.

We disregard relativistic corrections \cite{Barausse_2007, Speeney_2022} for the energy loss due to dynamical friction since they have only a small impact on the orbital evolution and thus the GW signal of a SMBHB here.

\subsubsection{Equations of motion \label{subsubsec:EoM}}
The coupled differential equations describing the secular evolution of the SMBHB parameters $e(t)$ and $a(t)$ can be formulated in an analogous way as in Refs.~\cite{Becker_2022, Cardoso_2021},
\begin{equation}
\begin{aligned}
     \frac{de}{dt} = \frac{e^2-1}{2 e} \left(\frac{dE_{\text{orb}}}{dt} \, \frac{1}{E_{\text{orb}}} + 2 \,\frac{dL_{\text{orb}}}{dt} \,\frac{1}{L_{\text{orb}}}\right) \; ,
\end{aligned}
\label{eq:de_dt}
\end{equation}
\begin{align}
    \frac{da}{dt} = \frac{dE_{\text{orb}}}{dt} \, \left(\frac{G M \mu}{2 a^2}\right)^{-1} \; ,
    \label{eq:da_dt}
\end{align}
\noindent where the total angular momentum $L_{\text{orb}}$ and total orbital energy $E_{\text{orb}}$ can be expressed as
\begin{align}
    L_{\text{orb}} = \mu r^2 \dot{\phi} = \mu \sqrt{G M a (1-e^2)} \; ,
    \label{eq:L_orb}
\end{align}
\vspace{-0.25cm}
\begin{align}
    E_{\text{orb}} = - \frac{G M \mu}{2 a} \; .
    \label{eq:E_orb}
\end{align}

We solve these equations numerically using our \texttt{SMBHBpy} code, which is publicly available at \url{https://github.com/DMGW-Goethe/SMBHBpy} (Ref.~\cite{Daniel_SMBHBpy}).

\subsection{GW signal of individual SMBHBs \label{subsec:gw-signal}}
The GW signal from eccentric SMBHBs is a superposition of multiple components, referred to as harmonics, occurring at integer multiples of the orbital frequency $f_\mathrm{orb}$ \cite{Peters_1963}. However, for small eccentricities, the binary system emits predominantly in the second harmonic \cite{Hamers_2021}.

Since we will only consider SMBHBs with circular orbits for later calculations, we focus on the equations for the two independent polarization modes $h_{+}$ and $h_{\cross}$ of the GW signals in the case of $e=0$ in the following. The equations for the GW signals with $e > 0$ can be found, for instance, in Ref.~\cite{Maggiore_V1_2007} and are also implemented in the \texttt{SMBHBpy} code.

The polarization modes for circular orbits are given by
\begin{align}
    h_{+}(t) = A(t)\, \frac{1 + \cos^2(\iota)}{2} \, \cos\left[\Phi(t)\right] 
    \; ,
    \label{eq:h_+}
\end{align}
\begin{align}
    h_{\times}(t) = A(t)\, \cos(\iota) \, \sin\left[\Phi(t)\right] \; ,
    \label{eq:h_x}
\end{align}
\noindent with the phase $\Phi(t)$  of the GW defined by
\begin{align}
    \Phi(t) = \int_{t_0}^{t} \omega(t') \, dt' \; ,
    \label{eq:GW_phase}
\end{align}
\noindent and 
\begin{align}
    A(t) = \frac{4\,\left[\left(1+z\right) G \mathcal{M}_c \right]^{5/3}}{D_L(z)\,c^4} (\pi f_{\text{obs}}(t))^{2/3} \; ,
    \label{eq:A(t)}
\end{align}
\noindent  where $\mathcal{M}_c = \mu^{3/5} M^{2/5}$ is the chirp mass and $D_L(z)$ is the luminosity distance depending on the redshift $z$. We compute $D_L(z)$ by assuming a $\Lambda$CDM universe \cite{Weinberg_1972} with cosmological parameters given by the Planck 2018 observations \cite{Planck_2020}. In addition, $f_{\text{obs}} = f/(1+z)$ is the observed GW frequency in the reference frame where the detector is located, $f$ is the corresponding GW frequency in the rest frame of the SMBHB, and $\iota$ denotes the inclination angle between the line-of-sight and the rotation axis of the binary system. The GW angular frequency $\omega(t)$ in Eq.~\ref{eq:GW_phase} corresponds to twice the orbital angular frequency $\omega_{\text{orb}}(t)$ of the SMBHB, i.e., $\omega(t) = 2\, \omega_{\text{orb}}(t)$ \cite{Maggiore_V1_2007}.

The strain $h(t)$, as the quantity of the GW signal that can be measured by a GW detector, is defined by the linear combination of $h_{+}$ and $h_{\cross}$
\begin{align}
    h(t) = F_{+} h_{+}(t) + F_{\times} h_{\times}(t)\;,
    \label{eq:h(t)}
\end{align}
\noindent where $F_{+,\times}$ are the detector pattern functions that depend in general on the observed GW frequency $f_{\mathrm{obs}}$ and the location of the SMBHB in the sky \cite{Coogan_2022}.

For simplicity, in the following analysis, we consider the case that the GW detector is optimally aligned for the $+$ mode. This is the case, for example, when $\iota = \pi/2$. In this case, $F_{+} = 1$ and $F_{\times} = 0$ \cite{Eda_2015}, which gives $h(t) = h_{+}(t) = A_{+}(t) \cos\left[\Phi(t)\right]$ with
\begin{align}
    A_{+}(t) = A(t) \, \frac{1 + \cos^2(\iota)}{2}\;.
    \label{eq:A_+}
\end{align}

It is useful to transition from the time $t$ representation of the GW signal to the frequency $f$ representation. The corresponding Fourier transformation of the GW signal is given by \cite{Eda_2015}
\begin{align}
    \tilde{h}(f)= \int_{-\infty}^{\infty} {h}(t) \,e^{2\pi i f t} \,dt\;.
    \label{eq:h_tilde_f}
\end{align}

To calculate $\tilde{h}(f)$, it is common to use the stationary phase approximation. This method neglects the rapidly oscillating GW phase $\Phi(t)$ and considers only the slowly oscillating amplitude $A_{+}(f)$ \cite{Eda_2015}. Within this approximation, the characteristic strain $h_c(f)$ is given by

\begin{align}
    h_c(f) = A_{+}(f)\,\sqrt{\frac{f^2}{\dot{f}}}\;.
    \label{eq:h_c(f)}
\end{align}

While $h_c(f)$, as defined above, captures the amplitude of the GW signal, it does not provide information about the GW phase evolution. To complement this, we introduce the dephasing, which quantifies the difference in the number of GW cycles, $N_\mathrm{cycles}$, accumulated between a binary system evolving in vacuum and in the presence of DM. The total number of cycles completed between an initial time $t_i$ and a final time $t_f$ is given by \cite{Becker_2022, Kavanagh_2020}

\begin{align}
N_\mathrm{cycles}(t_i, t_f) = \frac{1}{2\pi}\, \Phi(t_i, t_f) = \frac{1}{\pi}\int_{t_i}^{t_f} \omega_\mathrm{orb}(t)\,dt\;.
\label{eq:N_cycles}
\end{align}

Setting $t_f = t_\mathrm{ISCOs}$ as the time at which the SMBHB reaches $r_\mathrm{ISCOs} = r_\mathrm{ISCO,1} + r_\mathrm{ISCO,2}$, the dephasing can be calculated as

\begin{align}
\Delta N_\mathrm{cycles}(t) = N_\mathrm{cycles}^\mathrm{vacuum}(t, t_\mathrm{ISCOs}) - N_\mathrm{cycles}^\mathrm{DM}(t, t_\mathrm{ISCOs})\,.
\label{eq:DeltaN_cycles}
\end{align}

The dephasing can equivalently be evaluated in the frequency domain: $\Delta N_\mathrm{cycles}(t) \rightarrow \Delta N_\mathrm{cycles}(f_\mathrm{obs})$.

\subsection{Matter density as a function of $h_c$\label{subsec:rho(h_c)}}
In this subsection, we give a brief derivation for the matter density profile $\rho(r)$ around SMBHs as a function of the observed GW frequency $f_{\text{obs}}$ and the measured characteristic strain $h_c(f_{\text{obs}})$. We make the following assumptions:
\begin{itemize}
\item The orbits of $m_1$ and $m_2$ are circular, i.e., $e = 0$.
\item Both SMBHs have equal masses, $m_1 = m_2 = m$, so that $\rho_{1}(r) = \rho_{2}(r) = \rho(r)$ in Eq.~\ref{eq:dE_DF_} can be assumed.
\item There is only energy loss due to non-relativistic dynamical friction and GW emission.
\item GWs come into the detector from the optimal direction for the $+$ mode, i.e., $h(t) = h_{+}(t)$.
\end{itemize}

It should be noted that it is the first approximation that makes an analytic solution for $\rho(f_{\text{obs}})$ possible.

First, $\dot{\omega} = 2 \pi \dot{f}$ in Eq.~\ref{eq:h_c(f)} contains all information about the temporal evolution of the SMBHB and can be replaced by $\dot{\omega} = \partial \omega /\partial r \, dr/dt$ with
\begin{align}
    \frac{\partial \omega}{\partial r} = -3\, \sqrt{\frac{2 G m}{r^5}}\;.
    \label{eq:domega_dt}
\end{align}

By combining Eq.~\ref{eq:da_dt} with Eq.~\ref{eq:dE_orb}, the second factor can be rewritten as
\begin{align}
    \frac{dr}{dt} = \left[\left< \frac{dE_{\text{DF}}}{dt} \right> + \left< \frac{dE_{\text{GW}}}{dt} \right>\right] \left(\frac{G m^2}{2 r^2}\right)^{-1}\;.
    \label{eq:dr_dt_deri}
\end{align}

Under the above assumptions, $\left< dE_{\text{DF}}/dt\right>$ is directly proportional to $\rho(r)$. Substituting the final expression for $\dot{\omega}$ into Eq.~\ref{eq:h_c(f)}, the resulting equation can be solved for the matter density distribution. The result is 
\begin{align}
    \rho(f_{\text{obs}},h_c({f_{\text{obs}}})) = \mathcal{A} \, \frac{f^{10/3}_{\text{obs}}}{h^2_c(f_{\text{obs}})} - \mathcal{B}\, f_{\text{obs}}^{11/3}\;,
    \label{eq:rho_DM(h_c)}
\end{align}
\noindent with
\begin{align}
    \mathcal{A} = \frac{1}{3} \,\frac{\left[\pi^{4}  \, (1+z)^{16}\, G^{7} \, m^{10}\right]^{1/3} \, \left(1+\cos^2(\iota)\right)^2}{2^{{2/3}}\,\ln(\Lambda)\,D^2_L(z)\,c^8}\;,
    \label{eq:A_DM}
\end{align}

\begin{align}
    \mathcal{B} =\frac{2^{8/3}}{5}\, \frac{\left[\pi^{8} \,(1+z)^{11}\, G^{2}\, m^{5}\right]^{1/3}}{\ln(\Lambda)\,c^5}\;.
    \label{eq:B_DM}
\end{align}

To obtain Eq.~\ref{eq:rho_DM(h_c)}, we also used the correlation between $f_{\text{obs}}$ and $r$ given by
\begin{align}
    r = \left[\frac{2 G m}{\left(f_{\text{obs}}\, (1+z) \, \pi\right)^2}\right]^{1/3}.
    \label{eq:r_f_obs}
\end{align}

It should also be noted that $\rho(f_{\text{obs}})$ represents a radial density profile for any type of matter and is not limited to dark matter (model-independent).

\begin{figure*}[t]
    \centering
    \begin{subfigure}[t]{0.402\textwidth}
        \centering
        \includegraphics[width=\textwidth]{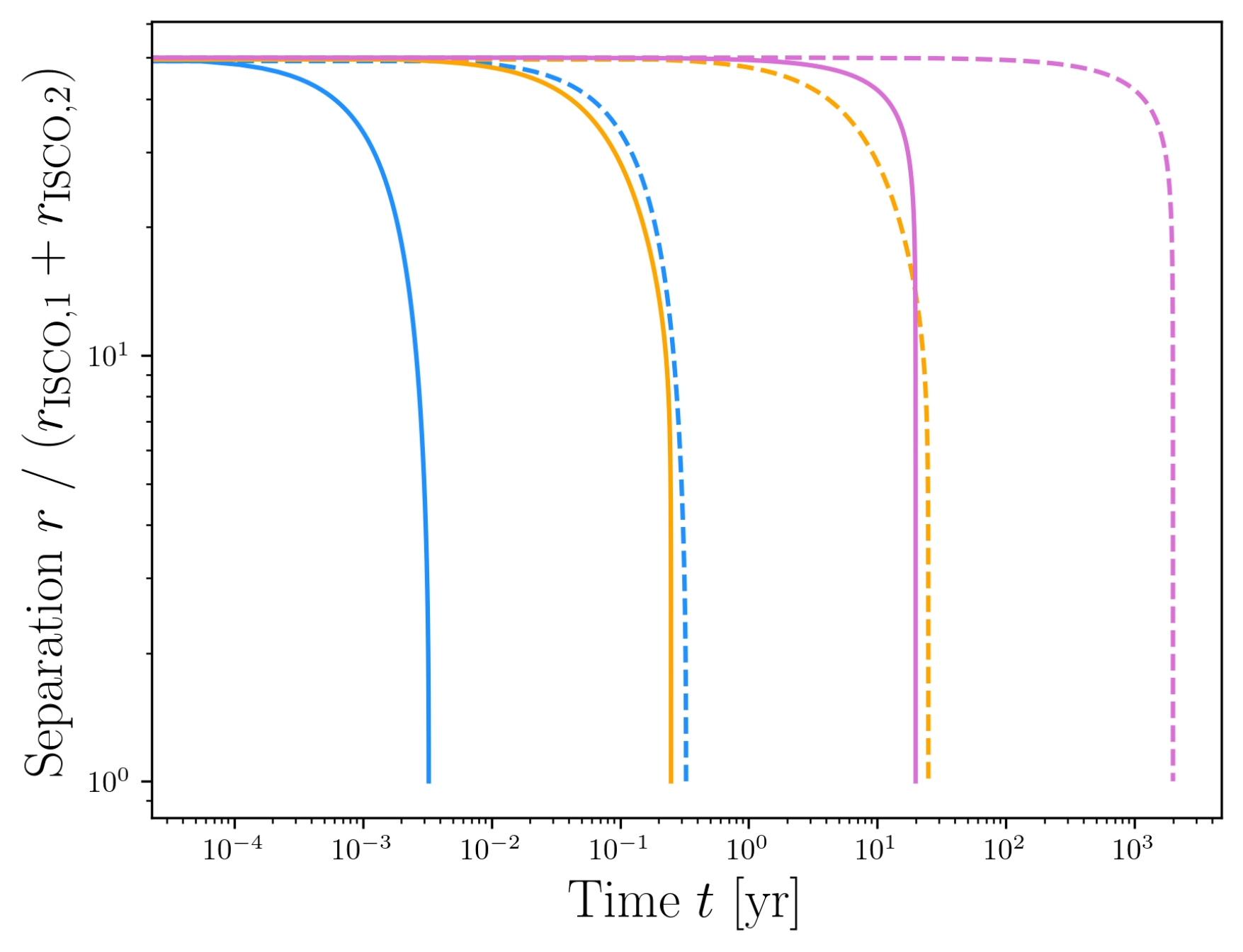}
    \end{subfigure}%
    \hfill
    \begin{subfigure}[t]{0.598\textwidth}
        \centering
        \raisebox{-0cm}{
            \includegraphics[width=\textwidth]{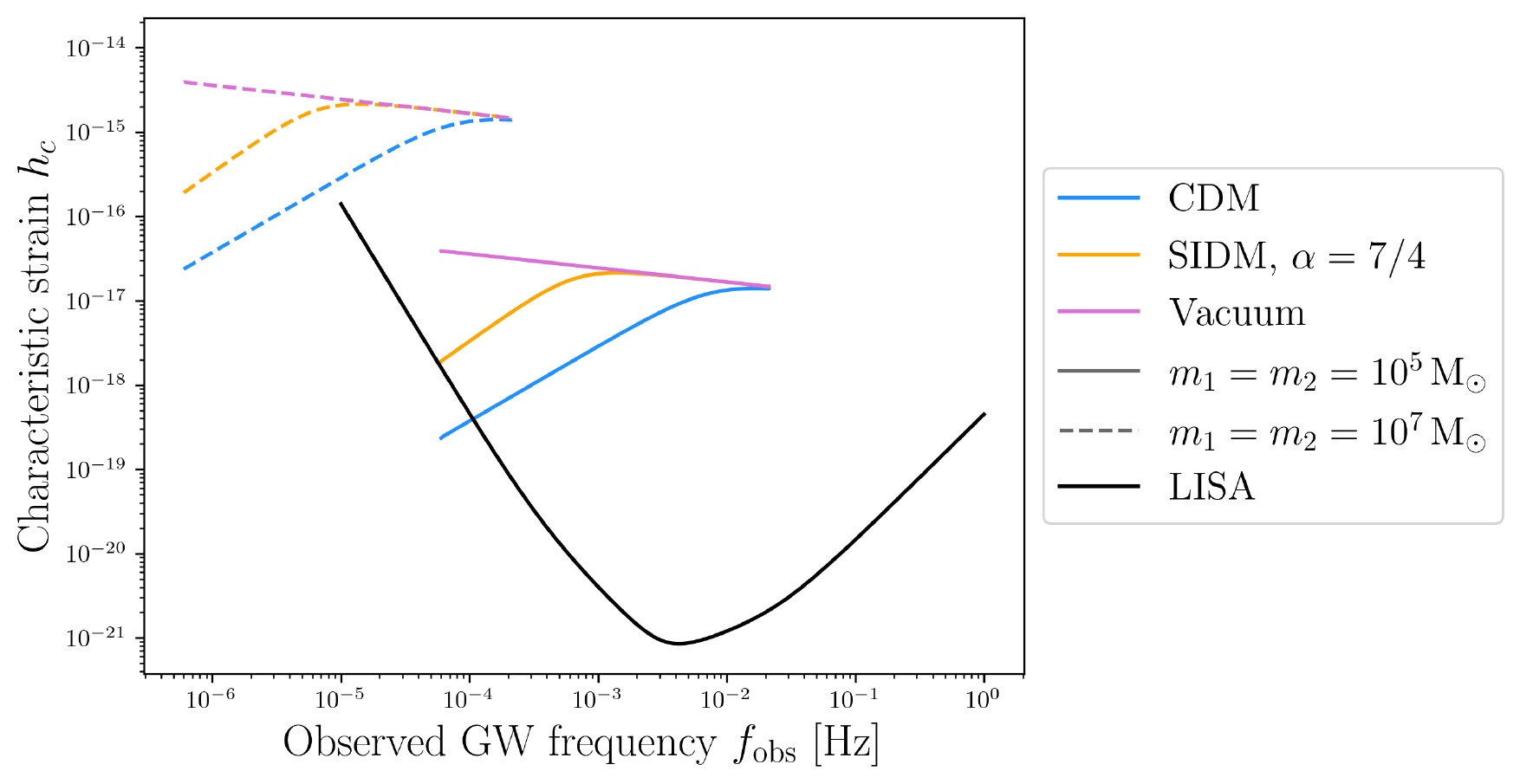}
        }
        \begin{minipage}{0.75\textwidth}
        \end{minipage}
    \end{subfigure}
    \caption{Comparison of the temporal evolution of two different equal-mass SMBHBs \textit{(left)} and their GW spectra \textit{(right)} for CDM (blue) and SIDM (orange) spikes with $\alpha = 7/4$. The case where the binary systems lose energy only through GW emission is represented by the pink lines (vacuum). \textbf{Solid lines}: For $m_1 = m_2 = 10^5\,\mathrm{M_{\odot}}$. \textbf{Dashed lines}: For $m_1 = m_2 = 10^7\,\mathrm{M_{\odot}}$. 
    \textit{Left:} Temporal evolution of the separation $r$ for circular orbits. Note that, in our setup, the initial separation is not fixed but grows with the mass of the SMBHB, resulting in an increase in the time until coalescence, even though the mass increases by two orders of magnitude. 
    \textit{Right:} GW spectra. The solid black line represents the projected sensitivity curve of LISA \cite{Robson_2019}.\newline
    \textit{Note:} For SIDM spikes with slopes smaller than $7/4$, the corresponding curves would lie closer to those for the vacuum case.}
    \label{fig:results_circ_orbits}
\end{figure*}

\section{Results \label{sec:results}}
In this section, we first investigate the influence of DM spikes on the GW signals from SMBHBs. We then consider how this effect will allow us to place constraints on the matter density around SMBHs using the LISA sensitivity curve.

\subsection{Impact of DM on GWs from SMBHBs \label{subsec:results_1}}
In the following, we consider the orbital evolution of two different equal-mass SMBHBs with $m = m_1 = m_2 = 10^5\,\mathrm{M_{\odot}}, \,10^7\,\mathrm{M_{\odot}}$ to provide a better understanding of how DM affects their orbits. To obtain the results presented in this subsection, we use the values for the respective parameters listed in Tab.~\ref{tab:1}.

\captionsetup[table]{name=TAB.}
\begin{table}[H]
\centering
\renewcommand{\arraystretch}{1.55}
\begin{tabular}{c|c}
Parameter & Value\\
\hline
\hline
$z$ & $0.03$ \\
\hline
$\sigma$ & $200\,\text{km/s}$ \\
\hline
$\ln(\Lambda)$ & $10$ \\
\hline
$\iota$ & $\pi/2$ \\
\hline
$r_0, \, a_0$ & $50\times (r_{\text{ISCO},1}+r_{\text{ISCO},2})$ \\
\hline
$a_\mathrm{SIDM}$ & $7/4$ ($b = 4$) \\
\end{tabular}
\centering
\vspace*{0.25cm}
\caption{Table with the values for the SMBHB parameters used for Figs.~\ref{fig:results_circ_orbits}-- \ref{fig:spectrogram_CDM_SIDM}.}
\label{tab:1}
\end{table}

In this subsection, we keep the other SMBHB and GW parameters fixed to better isolate the effects of the DM spikes. We assume a redshift of $z = 0.03$, corresponding to a luminosity distance of $D_L(z) \approx 136\,\mathrm{Mpc}$. $200\,\text{km/s}$ is a typical value for the velocity dispersion $\sigma$ in elliptical galaxies \cite{Forbes_1999}. We use this quantity to calculate the radius of the DM spikes (see Eq.~\ref{eq:r_h}). Additionally, we set the inclination angle $\iota = \pi/2$, so that $h(t) = h_{+}(t)$ (see \ref{subsec:gw-signal}). The parameters $r_0$ and $a_0$ denote the initial separation and semi-major axis, respectively. Their values are set to $50\times (r_{\text{ISCO},1}+r_{\text{ISCO},2})$ as an example. Furthermore, we choose $a_\mathrm{SIDM} = 7/4$ as our fiducial value, but also discuss how the SMBHB evolution and the associated GW signal differ for smaller SIDM spike slopes.

Before presenting our results, it is important to note that the energy deposited into the surrounding DM particles through dynamical friction can, in some cases, exceed the gravitational binding energy of a CDM spike, leading to its gradual saturation and dissipation over time \cite{Kavanagh_2020}. By closely following Ref.~\cite{Kavanagh_2020}, we estimate the ratio of these two energies for the SMBHBs considered here. We find that, under the main assumptions of circular orbits, equal-mass binaries, and spherically symmetric, static CDM spikes, the energy injected by the inspiralling binary is sufficient to unbind the spike rather quickly. The details of this estimate are presented in Appendix~\ref{sec:energy_balance_CDM}.

In contrast, as shown by Ref.~\cite{Alonso-Alvarez_2024}, SIDM may allow for partial rethermalization and replenishment of the spike, enabling such structures to survive for longer timescales. To qualitatively support this possibility, we provide a simple timescale estimate in Appendix~\ref{sec:timescale_estimate_SIDM}.

However, these conclusions must be interpreted with caution. The actual efficiency of spike depletion depends on several factors, including the orbital eccentricity, the spike slope, the merger history of the binary system, and the DM velocity distribution. In particular, Refs.~\cite{Mukherjee_2023, Kavanagh_2025} found, for intermediate-mass-ratio inspirals (IMRIs), that the disruption of a CDM spike proceeds significantly more slowly in N-body simulations than predicted by simplified analytic energy-balance arguments such as those adopted in Ref.~\cite{Kavanagh_2020} and in this work. This demonstrates that our conclusion -- namely, that CDM spikes around the SMBHBs studied here are unlikely to survive -- should be regarded as tentative.

We therefore emphasize that sophisticated numerical simulations of realistic SMBHB environments will be essential to fully assess the long-term stability of DM spikes.

In order to compare how different DM models influence the evolution of SMBHBs, we consider both CDM and SIDM spikes in this subsection. We note that the uncertainties in the spike stability discussed above do not impact the forecasted upper limits on the (dark) matter density presented in \ref{subsec:results_2}, as these bounds are model-independent.

\subsubsection{Circular inspirals \label{subsubsec:circ_inspiral}}
The evolution of circular orbits is shown in Fig.~\ref{fig:results_circ_orbits}, where the separation $r$ between the two SMBHs (expressed as a multiple of $r_{\text{ISCO},1}+r_{\text{ISCO},2}$) is plotted against the evolution time in years. Inspirals in the presence of DM spikes (blue and orange lines) proceed more rapidly than in the vacuum case (pink lines). Moreover, due to the higher central density of CDM spikes (see Fig.~\ref{fig:rho_vs_r_CDM+SIDM}), the inspiral is faster for both masses compared to the corresponding inspiral in SIDM spikes with $\alpha \leq 7/4$.

At first glance, one might conclude that for larger SMBH masses, the time until coalescence $t_\mathrm{c}$ increases. However, in our setup, the initial separation $r_0 = 50 \times (r_{\text{ISCO},1}+r_{\text{ISCO},2})$ scales with the SMBH mass $m$ according to $r_{\text{ISCO}} \propto m$ (see Eq.~\ref{eq:r_min}). This choice alone allows $t_\mathrm{c}$ to increase for higher masses.

\begin{figure}[t]
    \centering
    \includegraphics[width=0.98\columnwidth]{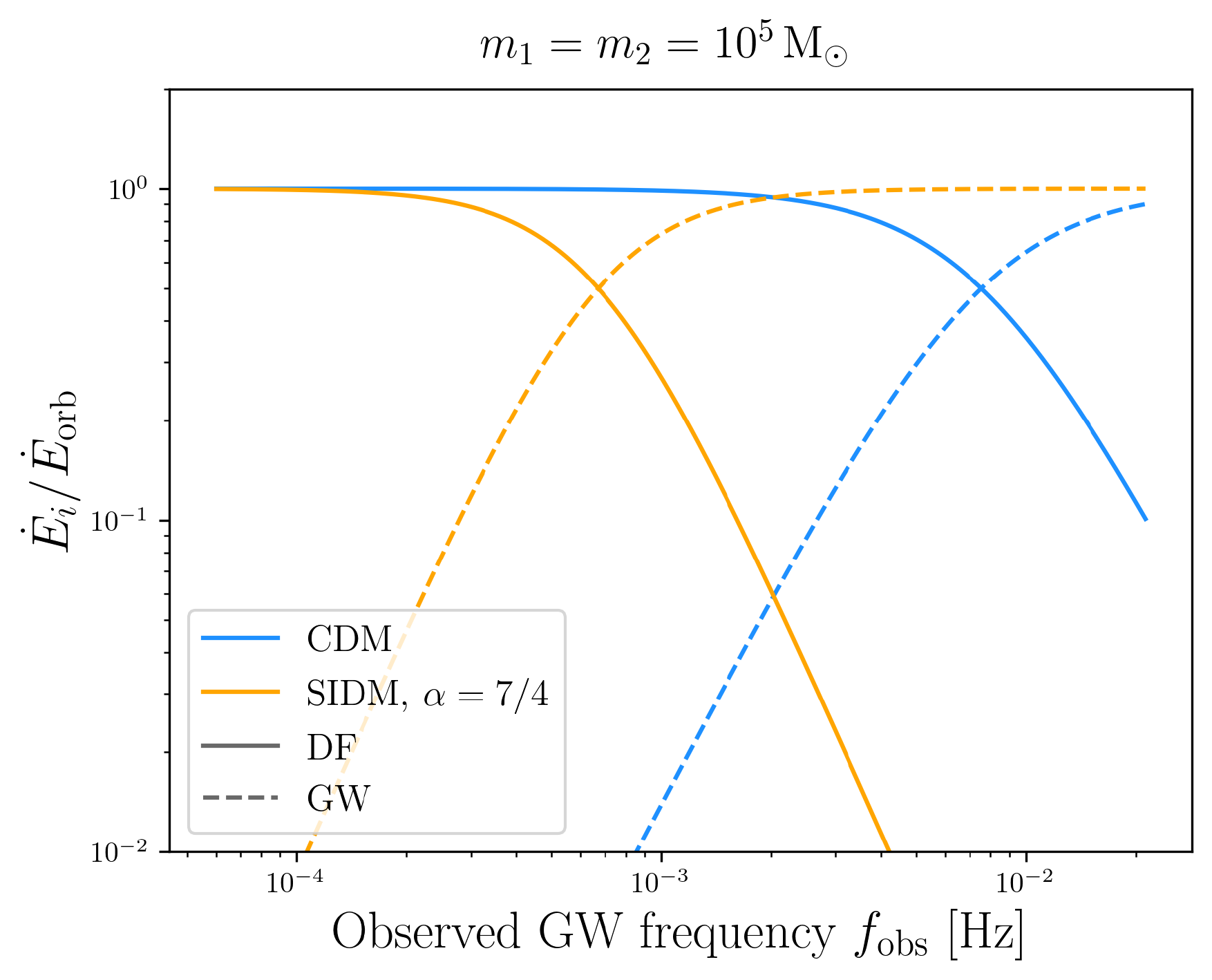}
    \caption{Temporal energy loss as a function of the observed GW frequency $f_\mathrm{obs}$ for the SMBHB with $m_1 = m_2 = 10^5\,\mathrm{M_\odot}$ and CDM (blue) or SIDM spikes with $\alpha = 7/4$ (orange). \textbf{Solid lines:} For dynamical friction ($i = \mathrm{DF}$). \textbf{Dashed lines:} For GW emission ($i = \mathrm{GW}$). \newline \textit{Note:} For $m_1 = m_2 = 10^7\,\mathrm{M_\odot}$ or SIDM spikes with smaller $\alpha$, the transition would occur at lower $f_\mathrm{obs}$.}
    \label{fig:energy_loss_vs_f_obs}
\end{figure}

\begin{figure}
    \centering
    \includegraphics[width=0.98\columnwidth]{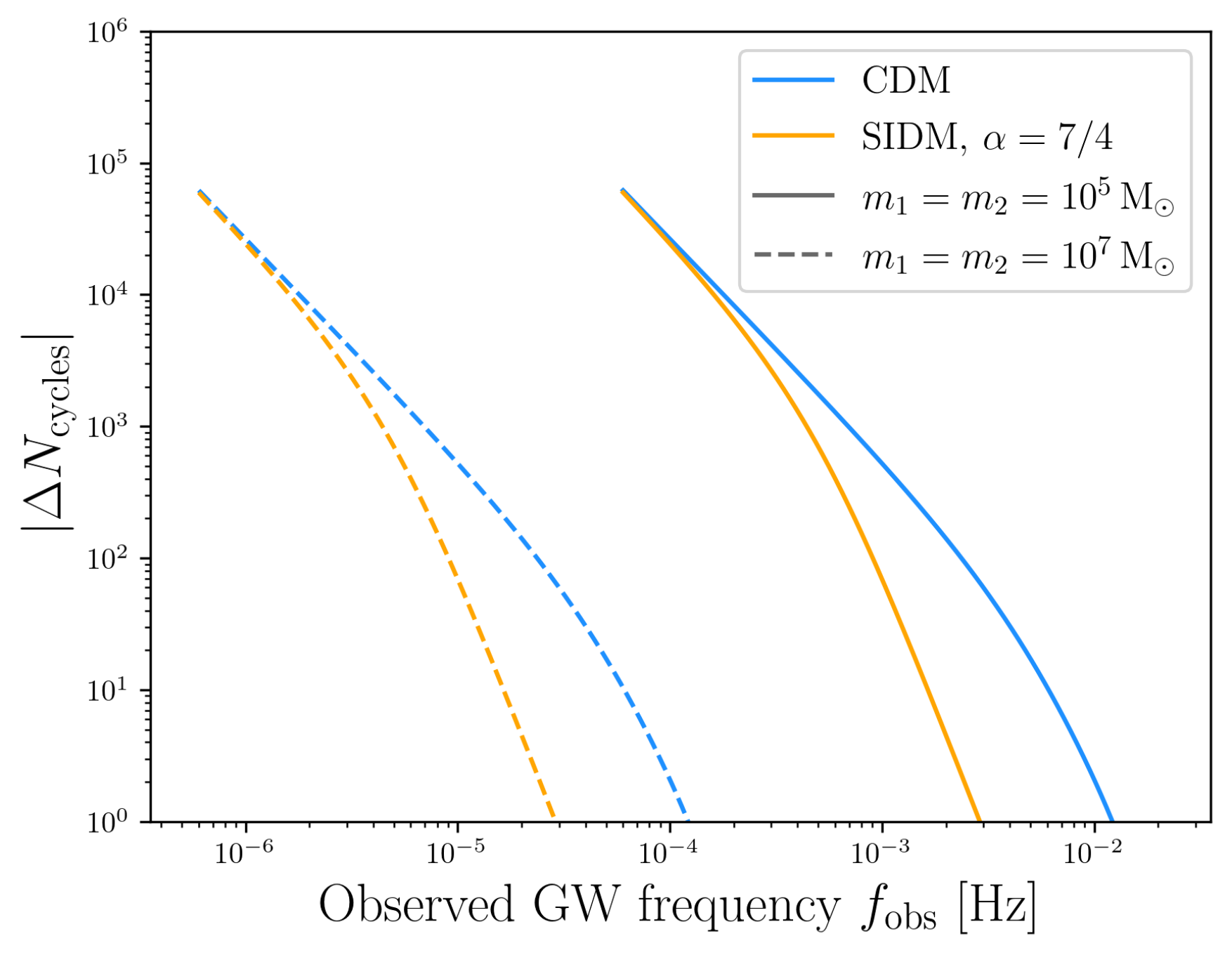}
    \caption{Dephasing as a function of $f_\mathrm{obs}$ for SMBHs surrounded by CDM (blue) or SIDM spikes with $\alpha = 7/4$ (orange). \textbf{Solid lines}: For $m_1 = m_2 = 10^5\,\mathrm{M_{\odot}}$. \textbf{Dashed lines}: For $m_1 = m_2 = 10^7\,\mathrm{M_{\odot}}$.\newline \textit{Note:} For SIDM spikes with $\alpha < 7/4$, the dephasing effect would be weaker across the full frequency range.}
    \label{fig:dephasing}
\end{figure}

\begin{figure*}[t]
    \centering
    \begin{subfigure}[t]{0.393\textwidth}
        \includegraphics[width=\linewidth]{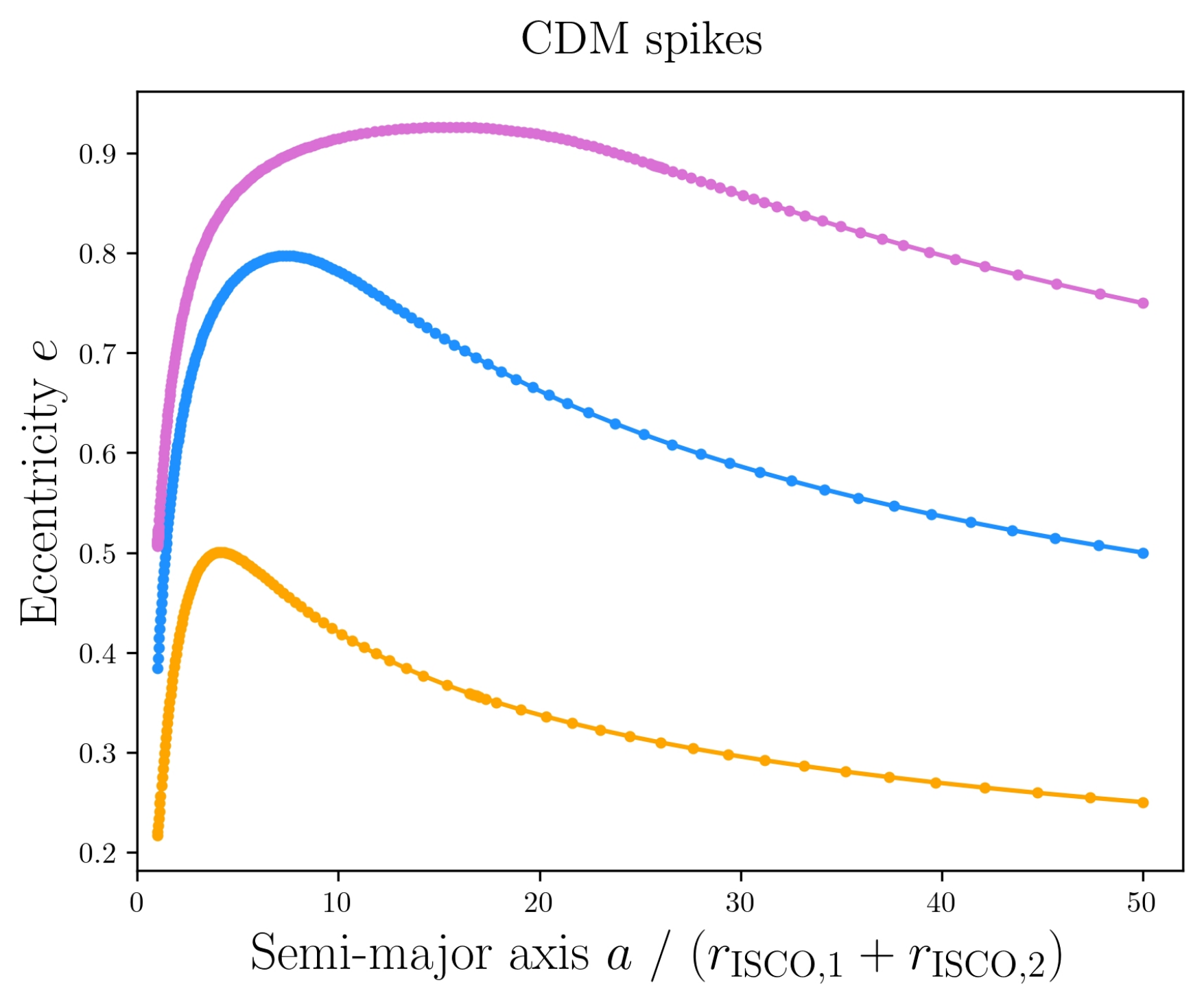}
    \end{subfigure}%
    \begin{subfigure}[t]{0.607\textwidth}
        \raisebox{0.005cm}{
            \includegraphics[width=0.95\textwidth]{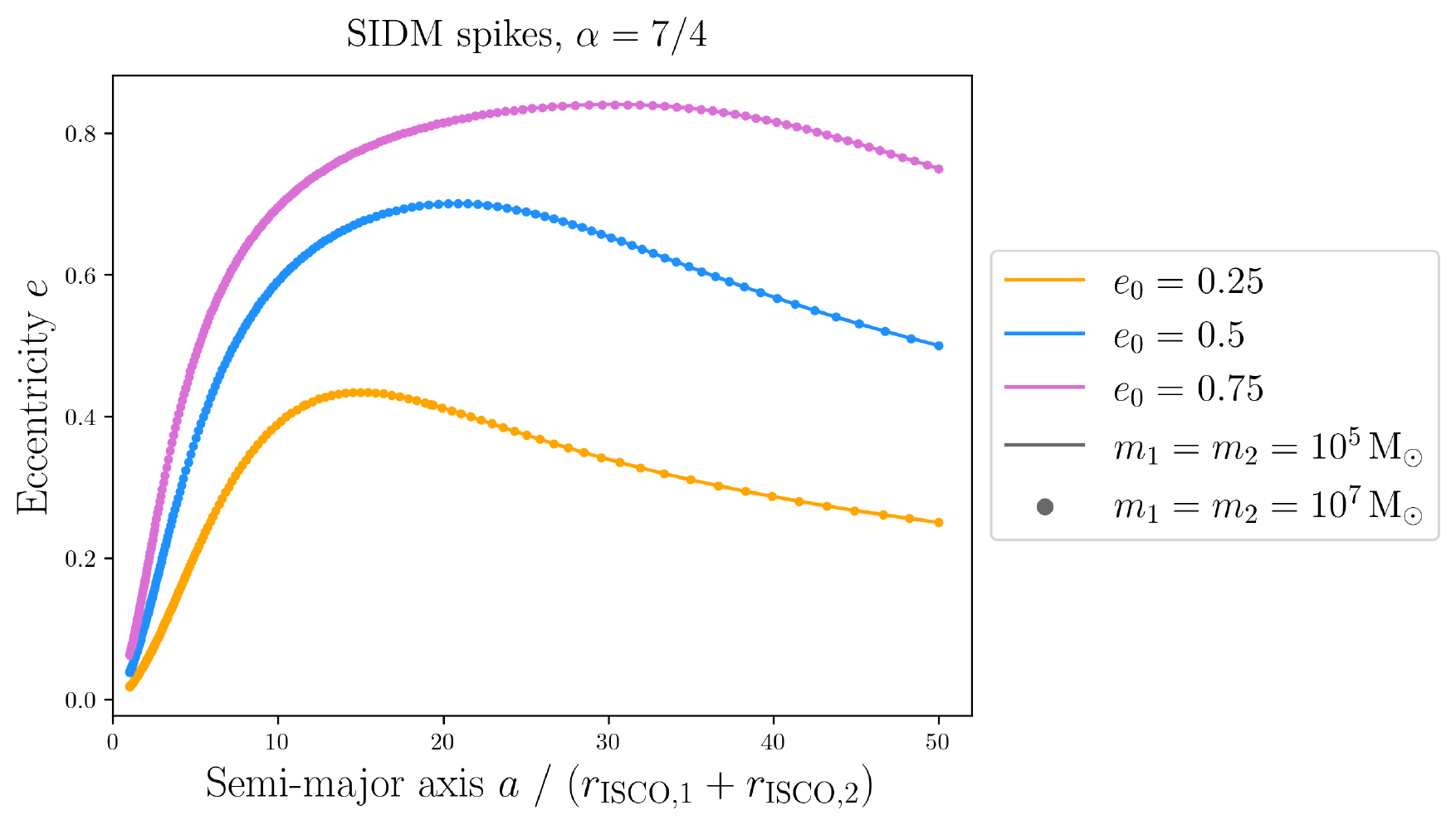}
        }
    \end{subfigure}
    \caption{Evolution of the eccentricity $e$ depending on the semi-major axis $a$ for different initial eccentricities $e_0 > 0$. For the left plot, CDM spikes were used, while for the right plot, SIDM with $\alpha = 7/4$ spikes were considered. \textbf{Solid lines}: For $m_1 = m_2 = 10^5\,\mathrm{M_{\odot}}$. \textbf{Dots}: For $m_1 = m_2 = 10^7\,\mathrm{M_{\odot}}$. The results are the same for both considered equal-mass SMBHBs (see text for explanation). \newline
    \textit{Note}: For the vacuum case, the two SMBHBs only lose energy through the emission of GWs. This effect circularizes the orbits throughout the entire evolution. Additionally, the smaller the SIDM spike slope $\alpha$, the weaker the eccentrification of the orbits due to dynamical friction.}
    \label{fig:eccentric_ev_e_a}
\end{figure*}

By comparing the GW spectra with and without DM spikes in Fig.~\ref{fig:results_circ_orbits}, it becomes clear that there are two phases in the evolution of SMBHBs, as mentioned in \ref{subsubsec:E-L-balance}. For the regime where the energy loss due to dynamical friction dominates, $h_c \propto f^{(5-\alpha)/3}_{\text{obs}}$ holds. Here, $\alpha$ represents the spike slope. The final phase, where GW emission dominates, is characterized by $h_c \propto f^{-1/6}_{\text{obs}}$. Fig.~\ref{fig:energy_loss_vs_f_obs} illustrates how the total energy loss is distributed among the different energy-dissipating mechanisms (dynamical friction: solid lines, GW emission: dashed lines) for a SMBHB with $m_1 = m_2 = 10^5\,\mathrm{M_\odot}$. This representation also provides an intuitive understanding for the positions of the turnovers, i.e., the frequencies where the slope of the GW spectrum changes. The turnovers are located at the transition between the two evolutionary phases, occurring around $7\times10^{-3}\,\mathrm{Hz}$ for CDM spikes (blue lines) and around $7\times10^{-4}\,\mathrm{Hz}$ for SIDM spikes with $\alpha = 7/4$ (orange lines). For SIDM spikes with $\alpha < 7/4$, the transition would take place at lower $f_\mathrm{obs}$, since the dynamical friction mechanism is less effective due to the reduced central density.

Additionally, Fig.~\ref{fig:results_circ_orbits} shows that for higher frequencies, especially for SIDM with $\alpha \leq 7/4$, $h_c$ cannot be distinguished from that of the vacuum case anymore. In this case, $h_c$ does not contain any information about the matter environment of the SMBHs. For CDM, a difference between the two GW spectra (with and without DM) can be observed up to the frequency corresponding to $r = r_{\text{ISCO},1} + r_{\text{ISCO},2}$ (here, the numerical calculation stops). Again, this can be explained by the higher central density of the CDM spikes. As a consequence, there are clear fingerprints of DM in the characteristic strain even at higher values of $f_{\text{obs}}$.

However, the impact of DM spikes on the evolution of SMBHBs can also be observed through the dephasing effect (see \ref{subsec:gw-signal} for details). In Fig.~\ref{fig:dephasing}, we compare the number of GW cycles completed in the presence and absence of CDM (blue) and SIDM with $\alpha = 7/4$ spikes (orange), respectively. Although the characteristic strain is significantly reduced over a certain frequency range due to dynamical friction with DM particles (see Fig.~\ref{fig:results_circ_orbits}), the corresponding dephasing in this regime is at a relatively high level. This effect becomes more pronounced with increasing DM density. Consequently, if the final inspiral of a SMBHB occurs within LISA’s sensitivity range, the impact of a high-density DM spike could still be clearly observable, despite a strong attenuation of the characteristic strain compared to the vacuum case, due to a substantial dephasing effect. Comparable results have also been presented for IMRIs in different DM distributions, e.g., in Refs.~\cite{Eda_2013, Eda_2015, Kavanagh_2020, Becker_2022, Kadota_2024}.

\subsubsection{Eccentric inspirals \label{subsubsec:ecc_inspiral}}
Although we specialize to circular inspirals for the rest of this paper, here we briefly examine the impact of DM spikes on eccentric Keplerian orbits.

The first thing to notice in Fig.~\ref{fig:eccentric_ev_e_a} is that the results for both masses (solid lines and dots) are the same for all initial eccentricities $e_0$. This is only the case because the components of the SMBHBs have the same mass and thus identical DM spike parameters, according to Eqs.~\ref{eq:rho_DM(r)}, \ref{eq:r_h}, \ref{eq:r_min} and \ref{eq:rho_sp}. At the same time, we assume $\ln(\Lambda) = 10$, which eliminates any implicit mass dependence in Eq.~\ref{eq:F_DF}. This behavior can also be easily shown by using $de/da = de/dt\,\left(da/dt\right)^{-1}$ and combining the equations that describe the evolution of the binary systems with DM spikes from \ref{subsec:orbitalevolution}.

Furthermore, the dynamical friction force $F_{\mathrm{DF}} \propto r^{-\alpha} v^{-2}$ leads to an eccentrification of the SMBH orbits (see Fig.~\ref{fig:eccentric_ev_e_a}). This is even generally the case for spike slopes $\alpha < 3$ as shown in the appendix of Ref.~\cite{Becker_2022}. On the other hand, the emission of GWs results in circularization. Additionally, $e$ converges to approximately zero during the evolution for all considered initial values $e_0$ in the case of SIDM spikes with $\alpha = 7/4$. This behavior is not seen for CDM spikes. In the latter scenario, the eccentricity converges to different final values for various $e_0$. Moreover, the maximum deviation of the eccentricity $e$ from its initial value $e_0$ is larger for CDM than for SIDM. This is due to the higher DM density around the SMBHs in the CDM model, which leads to a stronger dynamical friction force and a longer-lasting effect of eccentrification, i.e., up to a smaller $a$. Consequently, during the final phase, the time available is insufficient to achieve complete circularization by GW emission.

\begin{figure*}[t]
    \centering
    \begin{subfigure}[t]{0.393\textwidth}
        \includegraphics[width=\linewidth]{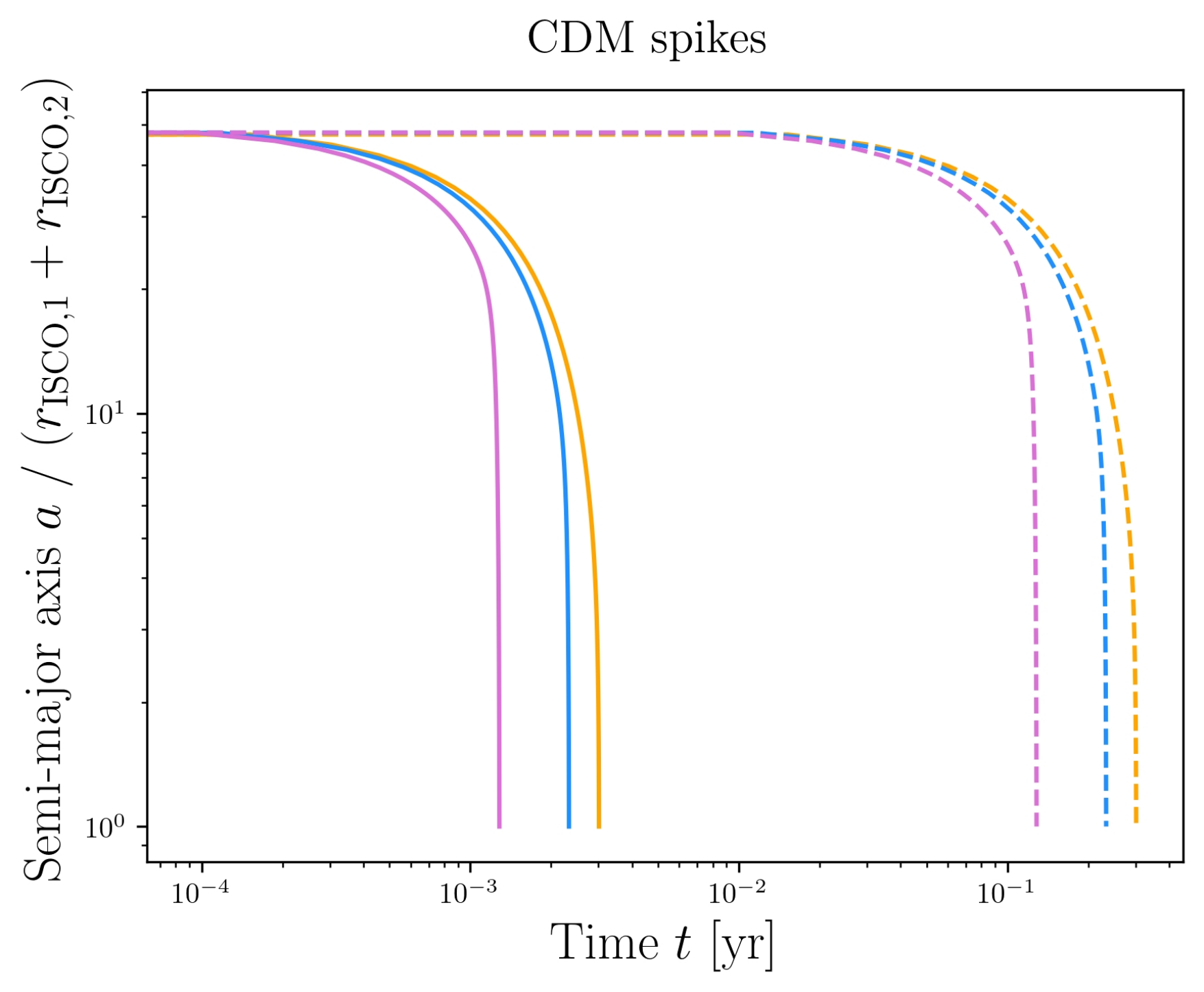}
    \end{subfigure}%
    \begin{subfigure}[t]{0.607\textwidth}
        \raisebox{-0cm}{
            \includegraphics[width=0.95\textwidth]{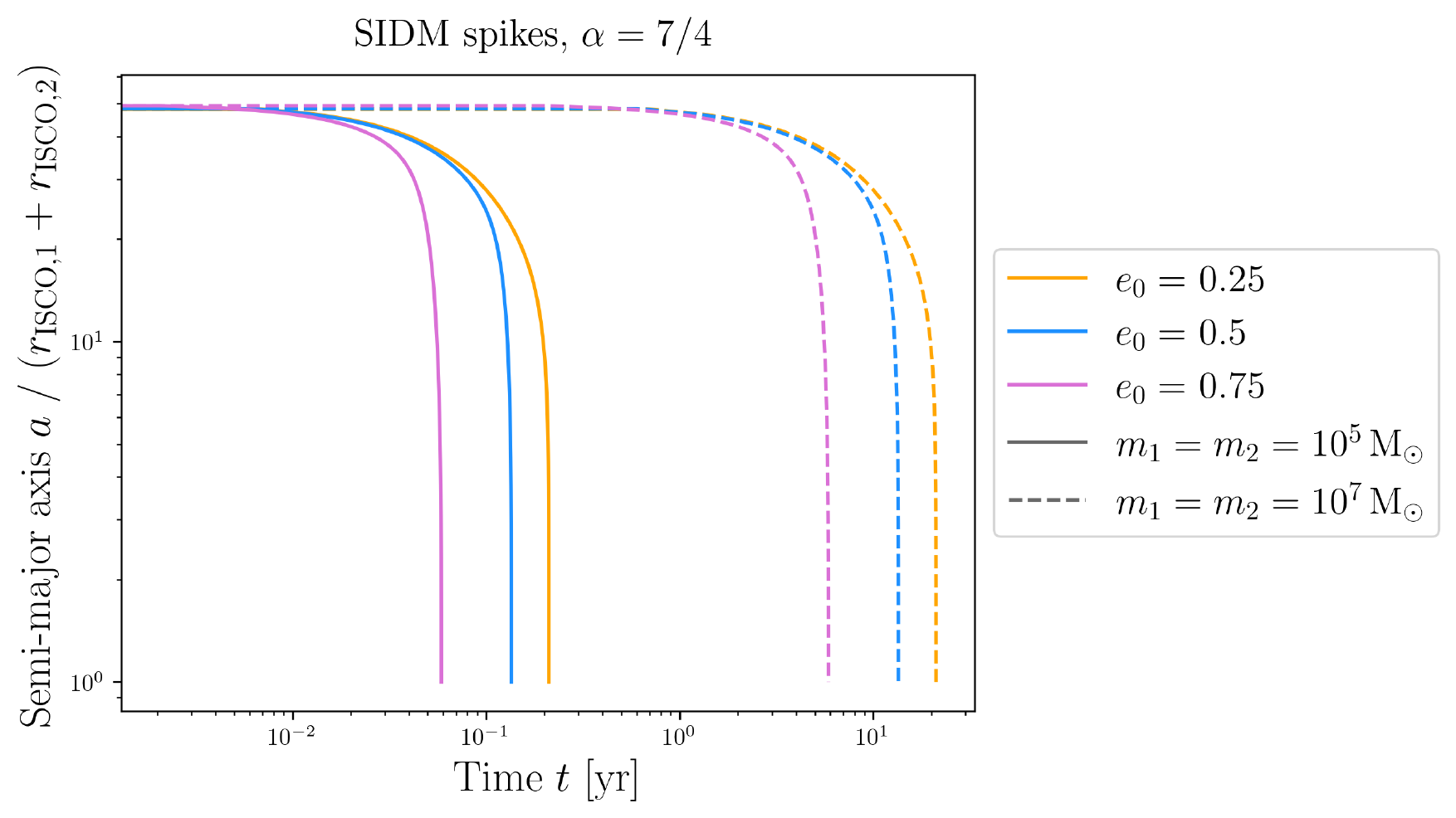}
        }
    \end{subfigure}
    \caption{Temporal evolution of the semi-major axis $a$ for different initial eccentricities $e_0 > 0$. For the left plot, CDM spikes were used, while for the right plot, SIDM spikes with $\alpha = 7/4$ were considered. \textbf{Solid lines}: For $m_1 = m_2 = 10^5\,\mathrm{M_{\odot}}$. \textbf{Dashed lines}: For $m_1 = m_2 = 10^7\,\mathrm{M_{\odot}}$.\newline 
    \textit{Note}: The results for the vacuum case would lie to the right of the depicted curves for each mass (see circular orbits from Fig. \ref{fig:results_circ_orbits}). For SIDM spikes with $\alpha < 7/4$, the binary evolution proceeds more slowly than in the case of $\alpha = 7/4$ for all considered $e_0$.}
    \label{fig:eccentric_ev_a_t}
\end{figure*}

The case of SIDM spikes with $\alpha = 7/4$ represents the scenario of strongest eccentricity growth due to dynamical friction among the SIDM profiles shown in Fig.~\ref{fig:rho_vs_r_CDM+SIDM}. Consequently, for those with $\alpha < 7/4$, the maximum eccentricity reached during the SMBHB evolution is smaller, and the phase dominated by GW emission extends to larger $a$.

In general, the temporal evolution of the semi-major axis $a$ shown in Fig.~\ref{fig:eccentric_ev_a_t} follows the same qualitative behavior as in the case of circular orbits. However, the higher the orbital eccentricity, the stronger the impact of DM dynamical friction becomes. This effect is more pronounced for CDM spikes owing to their higher central densities.

\begin{figure}[h]
    \centering
    \includegraphics[width=0.99\columnwidth]{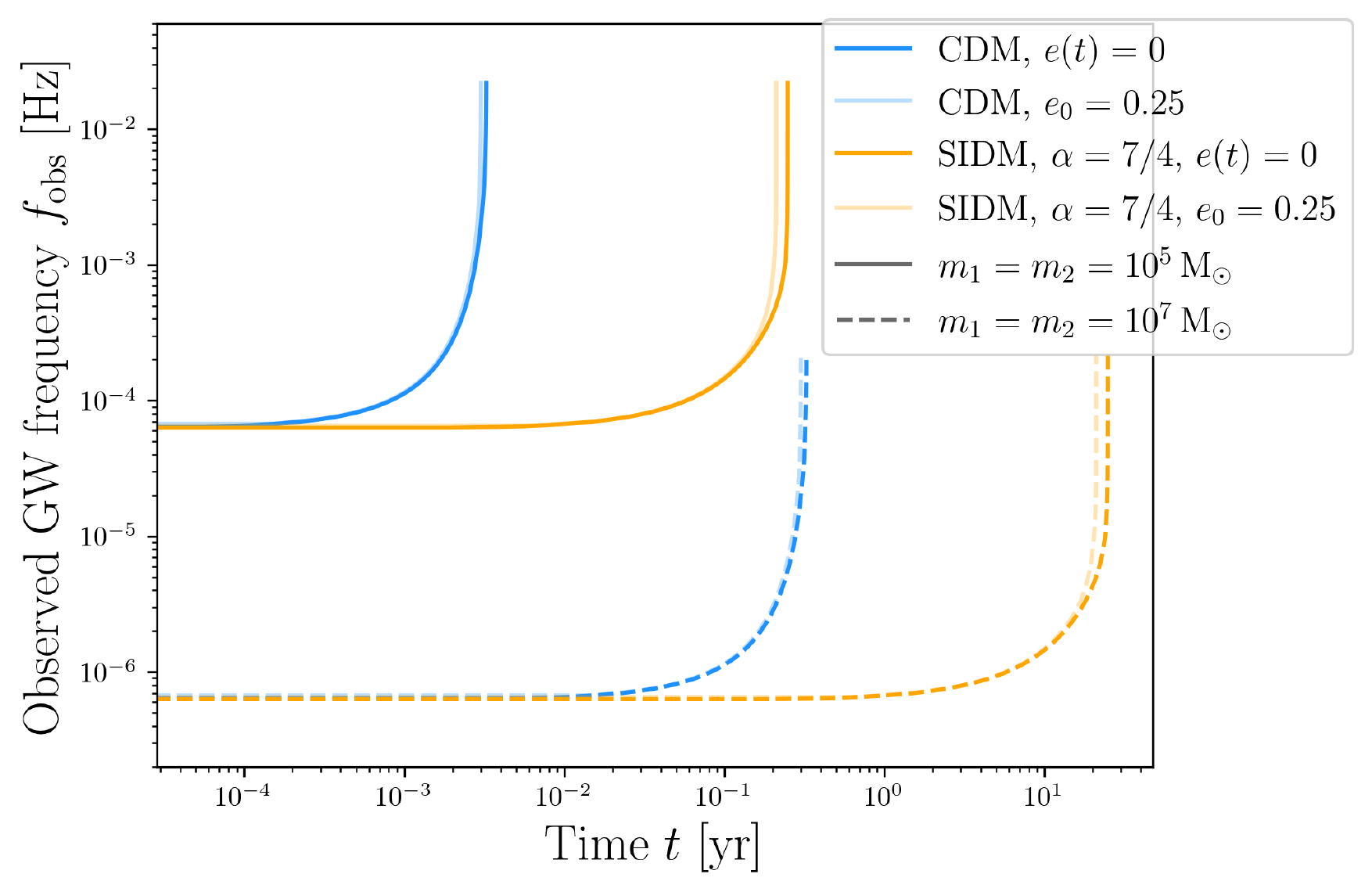}
    \caption{Spectrogram for SMBHs on circular ($e(t)=0$, opaque lines) and eccentric ($e_0 = 0.25$, faint lines) orbits, surrounded by CDM (blue) or SIDM spikes with $\alpha = 7/4$ (orange). \textbf{Solid lines}: For $m_1 = m_2 = 10^5\,\mathrm{M_{\odot}}$. \textbf{Dashed lines}: For $m_1 = m_2 = 10^7\,\mathrm{M_{\odot}}$. \newline \textit{Note:} For SIDM spikes with $\alpha < 7/4$, the orange curves would extend to later times.}
    \label{fig:spectrogram_CDM_SIDM}
\end{figure}

The acceleration of the SMBHB evolution for eccentric orbits compared to the circular case is evident in Fig.~\ref{fig:spectrogram_CDM_SIDM}, where the observed GW frequency $f_\mathrm{obs}$ is plotted as a function of time $t$ (spectrogram). We distinguish between the cases $e(t) = 0$ (opaque lines) and $e_0 = 0.25$ (faint lines), considering both CDM (blue lines) and SIDM spikes with $\alpha = 7/4$ (orange lines). In the eccentric case, we restrict the analysis to the second harmonic, meaning that -- as in the circular case -- $f_\mathrm{obs} = 2 f_\mathrm{orb}/(1+z)$ (see \ref{subsec:gw-signal}), where $f_\mathrm{orb}$ denotes the orbital frequency of the SMBHB.

The above findings suggest that SMBHBs may retain non-negligible orbital eccentricity, especially when the DM spike density is at a high level and the evolution starts with a large initial eccentricity. However, there are theoretical formulations for the dynamical friction force that result in the circularization of the orbits. This occurs when accounting for the fact that DM particles within the spikes move at different velocities relative to the SMBHs \cite{Kavanagh_2020, dosopoulou2023dynamical}. A recent study has further demonstrated that dynamical friction of DM particles moving faster than the secondary BH of eccentric IMRIs will eccentricify its orbit, independently of the spike slope $\alpha$ \cite{Zhou_2024_1}. Therefore, the question regarding the exact eccentricity evolution of SMBHBs with DM spikes remains unanswered and requires further investigation in the future.\\[0.1cm]
\indent In order to use the analytic formula for the matter density (Eq.~\ref{eq:rho_DM(h_c)}), we will only consider circular orbits in the following.

\begin{figure*}
    \centering
    \begin{subfigure}[t]{0.48\textwidth}
        \centering
        \includegraphics[width=\textwidth]{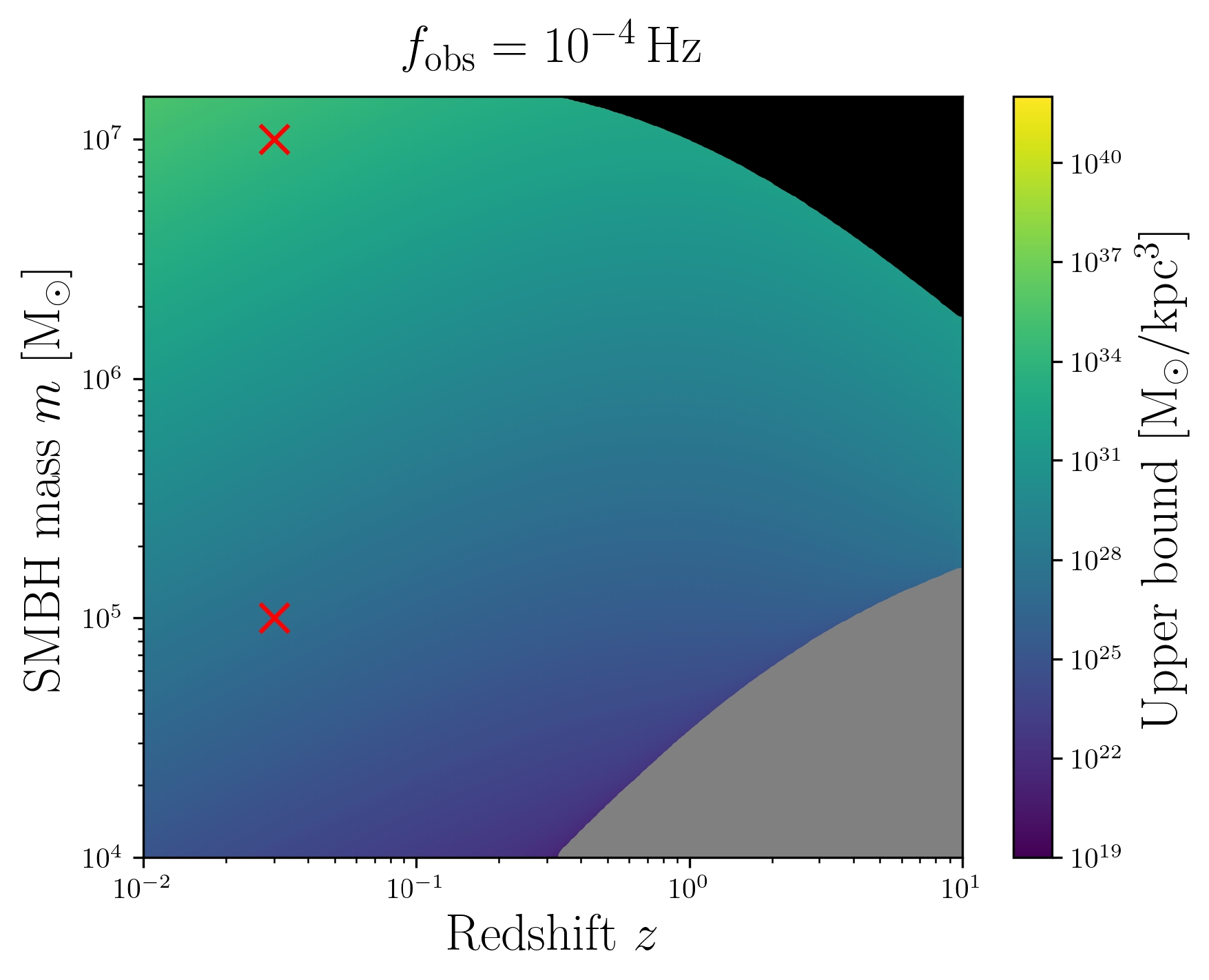}
        \caption{Forecasted upper bounds for different $m$ and $z$ at $f_{\mathrm{obs}} = 10^{-4}\,\mathrm{Hz}$.}
        \label{fig:LISA_thresholds_1e-4}
    \end{subfigure}%
    \hfill
    \begin{subfigure}[t]{0.48\textwidth}
        \centering
        \includegraphics[width=\textwidth]{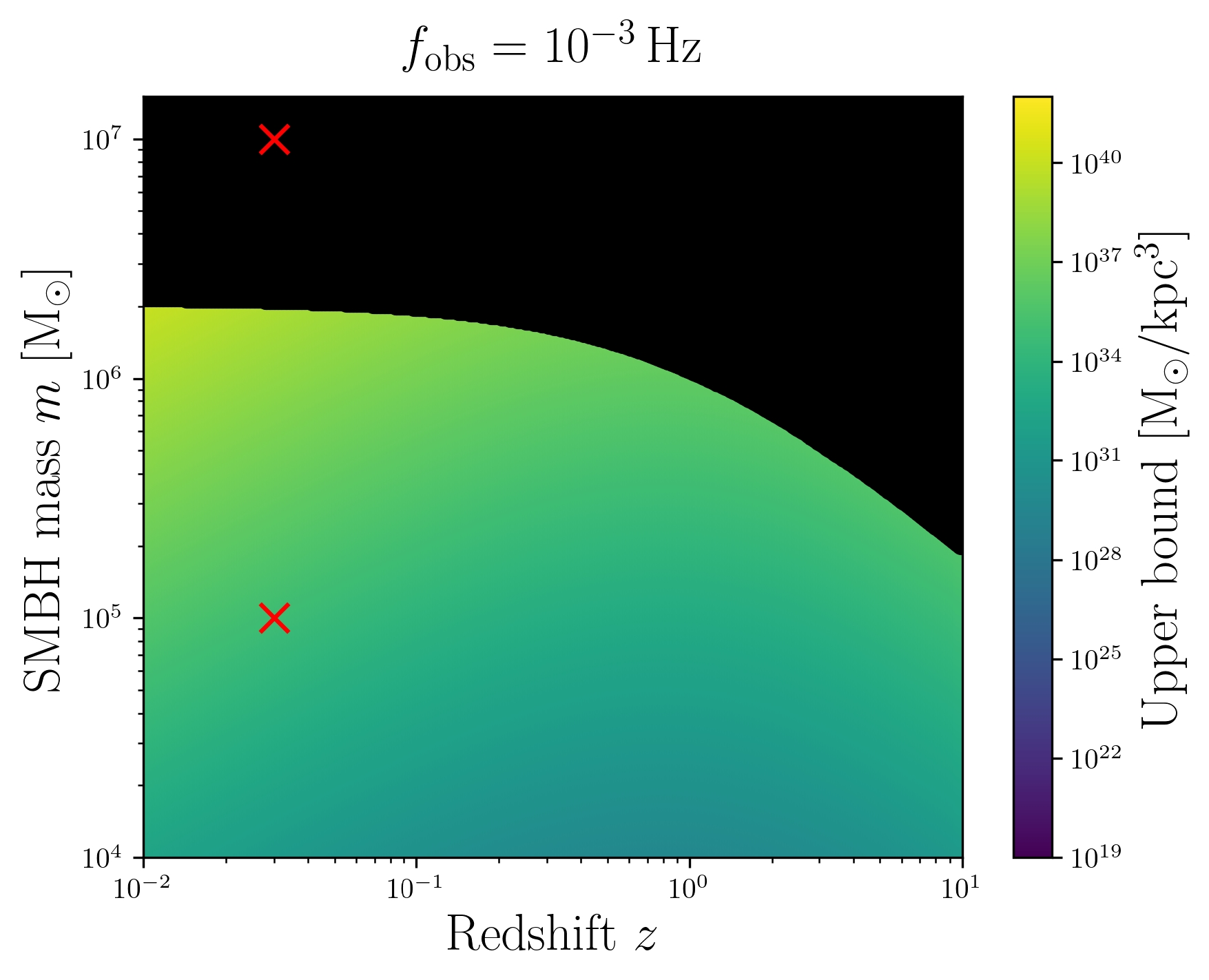}
        \begin{minipage}{\textwidth}
            \caption{Forecasted upper bounds for different $m$ and $z$ at $f_{\mathrm{obs}} = 10^{-3}\,\mathrm{Hz}$.}
        \label{fig:LISA_thresholds_1e-3}
        \end{minipage}
    \end{subfigure}
    \caption{Model-independent forecasted upper limits for the matter density around equal-mass SMBHs on circular orbits with different masses $m = m_1 = m_2$ and redshifts $z$ at two different $f_{\mathrm{obs}}$ within the LISA frequency range. The two red crosses mark the SMBHBs, which are examined separately in \ref{subsec:results_1} and in Fig.~\ref{fig:LISA_thresholds}. \textbf{Black area}: This parameter space is excluded since the corresponding GW frequencies $f_{\mathrm{ISCOs}}$ at $r_\mathrm{ISCO,1} + r_\mathrm{ISCO,2}$ would be smaller than $f_{\mathrm{obs}} = 10^{-4}\,\mathrm{Hz}$ and $f_{\mathrm{obs}} = 10^{-3}\,\mathrm{Hz}$, respectively. \textbf{Gray area}: Here, the characteristic strain of the corresponding GW signal lies already below the sensitivity curve of LISA at the respective $f_{\mathrm{obs}}$ in the vacuum case. $\{\ln(\Lambda)\,, \iota\} = \{10,\,\pi/2\}$.}
    \label{fig:LISA_thresholds_at_f_obs}
\end{figure*}

\subsection{Forecasted upper limits on the matter density for LISA \label{subsec:results_2}}
LISA will be able to detect GW signals from the last hours to years of the evolution of SMBHBs with masses between approximately $10^4\,\mathrm{M_{\odot}}$ and $10^7\,\mathrm{M_{\odot}}$ \cite{Amaro_Seoane_2023}. However, as can be seen from the GW spectra in Fig.~\ref{fig:results_circ_orbits}, the presence of matter in such systems leads to an attenuation of the characteristic strain compared to the vacuum case, which would affect the detectability of SMBHBs by LISA. The higher the matter density in the immediate vicinity of SMBHs, the more pronounced the difference in the GW signal with and without the influence of matter becomes in the final phase of their evolution. Consequently, for each SMBHB, there exists a maximum allowed matter density such that the corresponding GW signal can still be detected by LISA. 

Here, we consider equal-mass SMBHBs for a wide parameter space consisting of their masses $m = m_1 = m_2$ and redshifts $z$ and explore the possible upper limits that LISA could place on the matter density surrounding these binaries. We assume that the SMBHs are moving on circular orbits around their common center of mass, and that they lose energy through the dynamical friction mechanism as described in \ref{subsubsec:DF}. We then use the forecasted LISA sensitivity curve \cite{Robson_2019} and Eq.~\ref{eq:rho_DM(h_c)} to find the matter densities to which LISA would be sensitive. We extend our analysis down to binary systems with a signal-to-noise ratio (SNR) of 1, representing the threshold for detection. This approach allows us to establish upper bounds on the matter density around SMBHs based on the minimum SNR detectable by LISA.

In Fig.~\ref{fig:LISA_thresholds_at_f_obs}, we present the derived model-independent upper bounds on the (dark) matter density in SMBHBs with BH masses $m$ between $10^4\,\mathrm{M_{\odot}}$ and $10^7\,\mathrm{M_{\odot}}$ at redshifts $z$ ranging from $0.01$ to $10$ for two different GW frequencies $f_\mathrm{obs} = 10^{-4}, 10^{-3}\,\mathrm{Hz}$. Different colors represent different values for the upper limit on the density. Additionally, two regions within the considered parameter space are ruled out from the outset. For binary systems with masses and redshifts within the gray area, the GW signal at the indicated $f_\mathrm{obs}$ lies below the LISA sensitivity curve when the SMBHs evolve in vacuum. Therefore, these circular SMBHBs are generally not detectable by LISA at this $f_\mathrm{obs}$. For parameter combinations in the black region, the GW frequency $f_\mathrm{ISCOs}$ at the separation $r_\mathrm{ISCO,1} + r_\mathrm{ISCO,2}$ between the SMBHs of the corresponding binary system is smaller than the fixed value for $f_{\mathrm{obs}}$. However, for $f_{\mathrm{obs}} > f_{\mathrm{ISCOs}}$, the Newtonian approximation employed in this work is no longer valid. Therefore, no conclusions can be drawn about the matter densities in this region.

\begin{figure*}[t]
    \centering
    \begin{subfigure}[t]{0.399\textwidth}
        \centering
        \includegraphics[width=\textwidth]{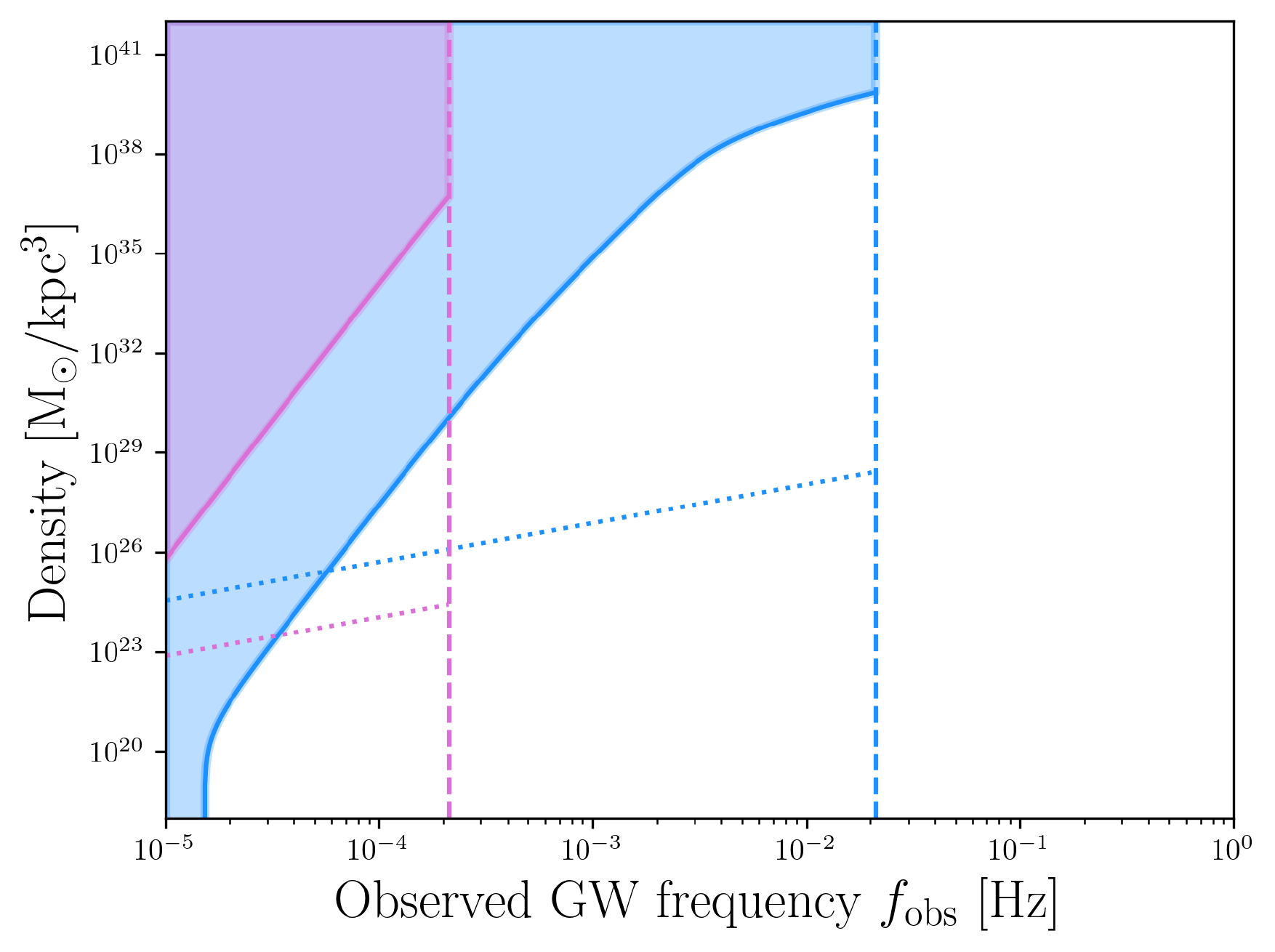}
        \caption{Forecasted upper limits as a function of the observed GW frequency $f_{\mathrm{obs}}$.}
        \label{fig:LISA_thresholds_f}
    \end{subfigure}%
    \hfill
    \begin{subfigure}[t]{0.601\textwidth}
        \centering
        \raisebox{-0.015cm}{
            \includegraphics[width=0.96\textwidth]{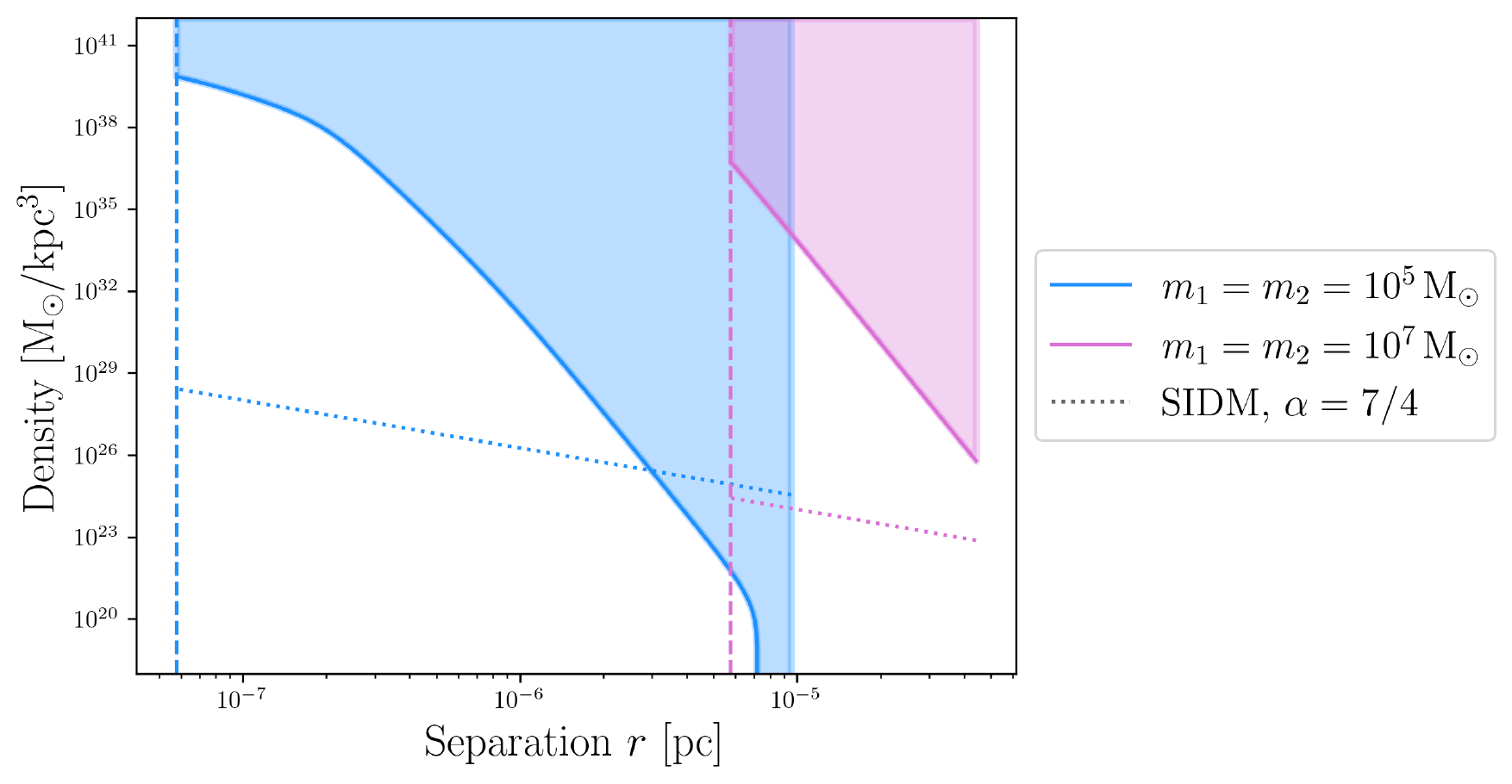}
        }
        \begin{minipage}{0.75\textwidth}
        \vspace{-0.175cm}
            \caption{Forecasted upper limits as a function of the separation $r$.}
        \label{fig:LISA_thresholds_r}
        \end{minipage}
    \end{subfigure}
    \caption{\textbf{Solid lines}: Model-independent forecasted upper limits for the matter density in two selected SMBHBs with circular orbits that exhibit masses of $m_1 = m_2 = 10^5\,\mathrm{M_{\odot}}, 10^7\,\mathrm{M_{\odot}}$ (blue, pink) and a redshift of $z = 0.03$. \textbf{Dotted lines}: Density distribution for the corresponding SIDM spikes with $\alpha = 7/4$. \newline
    The sum of the ISCOs for each SMBHB under consideration is represented by a vertical dashed line. $\{\sigma,\, \ln(\Lambda)\,, \iota\} = \{200\,\text{km/s},\,10,\,\pi/2\}$.\newline
    \textit{Note:} We use the specified value for the velocity dispersion $\sigma$ to calculate the SIDM spike radius, according to Eq.~\ref{eq:r_h}.}
    \label{fig:LISA_thresholds}
\end{figure*}

Moreover, Fig.~\ref{fig:LISA_thresholds_at_f_obs} demonstrates that the upper limit increases with increasing mass $m$ for a fixed GW frequency $f_{\mathrm{obs}}$ and redshift $z$. This is because more massive SMBHBs produce a stronger GW signal (see Fig.~\ref{fig:results_circ_orbits}), which lies further above LISA's sensitivity curve compared to that of lighter binary systems. Consequently, the attenuation of the characteristic strain due to the presence of matter in these SMBHBs can be greater, while still allowing the detection by LISA. This also explains why the upper bounds are larger at smaller redshifts compared to higher values for $z$ for a fixed BH mass $m$.

In the following, we will study the allowed values for the matter density across the entire frequency range of LISA for the two SMBHBs, whose temporal evolution and GW signals have already been discussed in the previous subsection. These two binary systems are also marked with red crosses in Fig.~\ref{fig:LISA_thresholds_at_f_obs}.

The corresponding results for these two selected equal-mass SMBHBs with $z = 0.03$ and $m=  m_1 = m_2 = 10^5\,\mathrm{M_{\odot}}, 10^7\,\mathrm{M_{\odot}}$ are depicted in Fig.~\ref{fig:LISA_thresholds} as a function of the observed GW frequency $f_{\mathrm{obs}}$, and the separation $r$ between the SMBHs. As a reference, the density distributions for SIDM spikes with $\alpha = 7/4$ are plotted, along with the sum of the ISCOs of the considered SMBHs.

Fig.~\ref{fig:LISA_thresholds_r} shows that the density of SIDM spikes decreases with increasing SMBH mass. This is not surprising, as according to Eqs.~\ref{eq:rho_DM(r)}, \ref{eq:r_h}, and \ref{eq:rho_sp}, it holds that $\rho_{\text{DM}}(r) \propto m^{\alpha-2}$, where $\alpha$ is the spike slope, meaning $\rho_{\text{SIDM}}(r) \propto m^{- 1/4}$ in the case of $\alpha = 7/4$. Therefore, the same trend is anticipated for $\alpha < 7/4$.

Furthermore, it can be seen in Fig.~\ref{fig:LISA_thresholds} that the forecasted upper limit for the matter density increases with increasing mass over the entire restricted separation range. By combining Eqs.~\ref{eq:rho_DM(h_c)} and \ref{eq:r_f_obs}, we obtain the following relation between the mass and the maximum allowed matter density $\rho_{\mathrm{c}}$ as a function of $r$: $\rho_{\mathrm{c}}(r) \propto \left(m^5-m^{7/2}\right)$.

For matter densities in the blue/pink filled areas, the GW signal is below the LISA sensitivity curve and is therefore not observable. Thus, for example, the effects of SIDM spikes with $\alpha = 7/4$ on the characteristic strain of the binary system with $m = 10^7\,M_{\odot}$ would already be visible from the lower end of the LISA frequency range (about $10^{-5}\,\mathrm{Hz}$), whereas the influence of the spikes around the SMBHs with $m = 10^5\,M_{\odot}$ would only become noticeable from around $10^{-4}$ Hz. This aligns with the results from Fig.~\ref{fig:results_circ_orbits}. 

\begin{figure}[t]
    \centering
    \includegraphics[width=0.99\columnwidth]{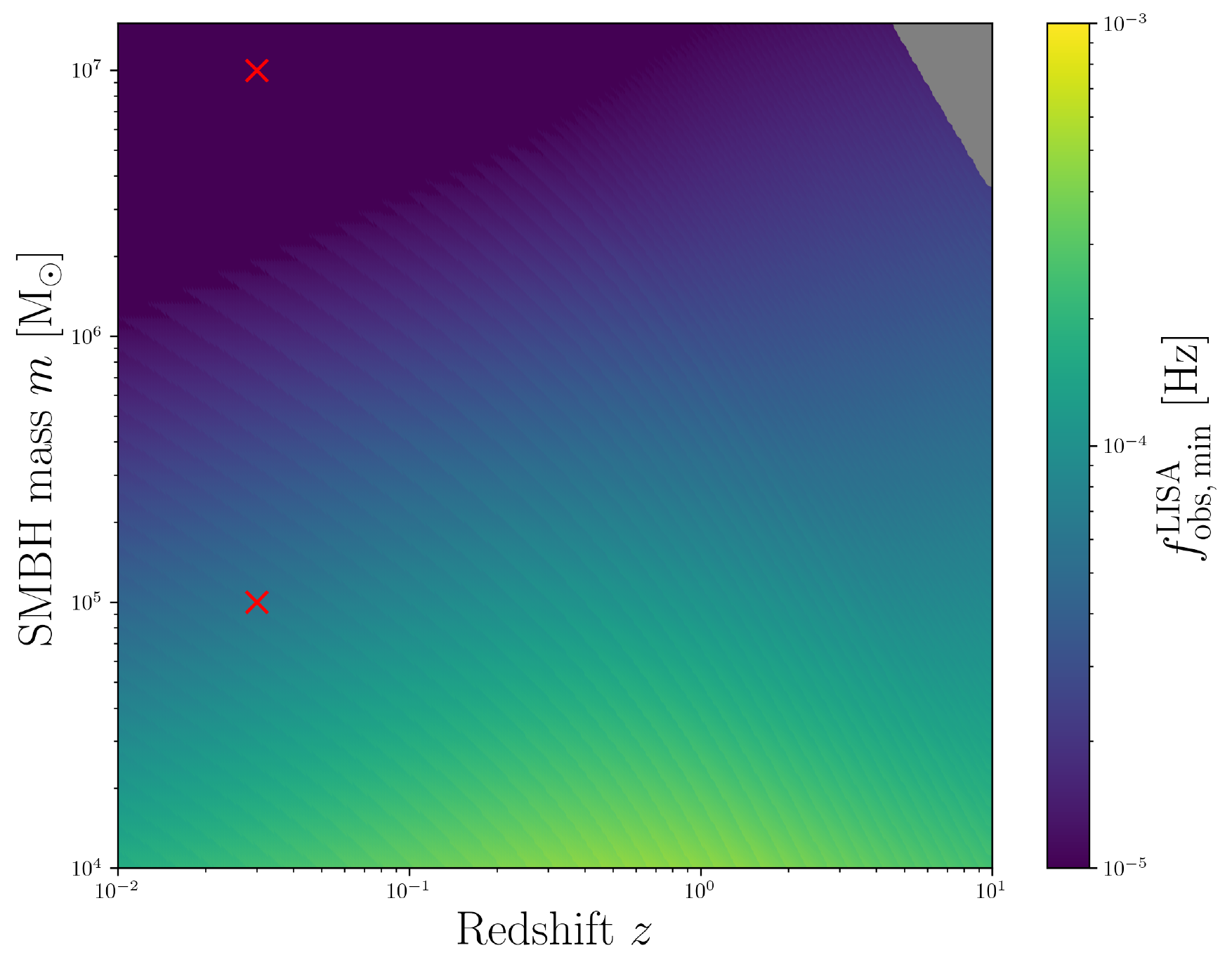}
    \caption{Observed frequency $f^{\mathrm{LISA}}_{\mathrm{obs,min}}$ from which the GW signal of circular, equal-mass SMBHBs with varying $m$ and $z$ becomes detectable by LISA ($\mathrm{SNR} \geq 1$). Each binary component is assumed to have a SIDM spike with $\alpha = 7/4$. The two red crosses mark the SMBHBs, which are examined separately in \ref{subsec:results_1} and in Fig.~\ref{fig:LISA_thresholds}. \textbf{Gray area:} Here, the binary systems evolve entirely outside LISA’s observational reach. $\{\sigma,\, \ln(\Lambda)\,, \iota\} = \{200\,\text{km/s},\,10,\,\pi/2\}$}
    \label{fig:f_min_LISA}
\end{figure}

To generalize this finding beyond the two specific examples discussed above, we compute the minimum observed frequency  $f^{\mathrm{LISA}}_{\mathrm{obs,min}}$, above which LISA is sensitive ($\mathrm{SNR} \geq 1$) to the GW signal from circular, equal-mass SMBHBs with varying component mass $m$ and redshift $z$, assuming that each SMBH is surrounded by a SIDM spike with $\alpha = 7/4$. The results are shown in Fig.~\ref{fig:f_min_LISA}. The same parameter space for $m$ and $z$ as in Fig.~\ref{fig:LISA_thresholds_at_f_obs} is used. SMBHBs with parameters lying in the gray region evolve entirely outside LISA’s observational reach. We note that the shape of LISA’s sensitivity curve strongly influences the distribution of $f^{\mathrm{LISA}}_{\mathrm{obs,min}}$ across the explored parameter space. For instance, $f^{\mathrm{LISA}}_{\mathrm{obs,min}}$ decreases with increasing $m$ for a fixed redshift $z$.

In summary, our results show the frequency range within which a SMBHB with a specific density distribution will be observable by LISA in the future. Depending on the mass $m$ and redshift $z$ of the binary system, there exists a forecasted upper limit for the (dark) matter density. This fact limits the number of SMBHBs that can be studied using LISA. However, it is reassuring to note that the observation threshold regarding the matter density for all considered SMBHBs is at a high level. Some of these densities seem to be unphysical, but when we compare them with the expected densities of neutron stars of $\mathcal{O}(10^{45}\,\mathrm{M_{\odot}}/\mathrm{kpc^3})$ \cite{Abbott_2018}, they are high but still within an acceptable range. Observing binary systems with LISA below these upper bounds could enhance our understanding of the particle properties of the surrounding matter and the orbital evolution of SMBHs based on the dynamical friction mechanism.

\section{Conclusions\label{sec:conclusions}}
In this study, we first investigate the influence of DM spikes around SMBHs in binary systems on their orbital evolution, and the GW signals they emit within the frequency range accessible to LISA. Our results demonstrate that both the presence of matter in a SMBHB and eccentric orbits can significantly reduce the time until coalescence. Our description of dynamical friction leads to an eccentrification of the SMBH orbits, as also discussed in Ref.~\cite{Cardoso_2021}. Therefore, high DM densities ensure that the remaining time within the GW emission phase is insufficient to fully circularize orbits with large initial eccentricities. Hence, it appears plausible to detect SMBHBs that exhibit an intermediate eccentricity during the late stage of their evolution. 

We show that the presence of surrounding matter in circular SMBHBs leads to an attenuation of the emitted GW signal compared to the vacuum case. Building on these findings, we develop a model-independent approach to compute an ``observation threshold'' for the matter density distribution in a SMBHB from a given limit on the characteristic strain and tested it using LISA's sensitivity curve. Our analysis shows that while the upper bounds are high, they are still below the expected densities within neutron stars of $\mathcal{O}(10^{45}\,\mathrm{M_\odot}/\mathrm{kpc^3})$ \cite{Abbott_2018}. Consequently, LISA should have the potential to measure GWs from a variety of SMBHBs with matter densities over several orders of magnitude. The expected densities for SIDM spikes with $\alpha \leq 7/4$ around SMBHs are within this range.

If LISA detects GWs from an individual SMBHB in the final phase of its evolution and if this binary system satisfies the conditions outlined in \ref{subsec:rho(h_c)}, one can calculate the matter density in its environment using Eq.~\ref{eq:rho_DM(h_c)}. By comparing the result with the density distributions expected for different models, possible constraints on the matter surrounding the SMBHs and on its physical properties can be derived.

We also emphasize that information about the DM environment of a SMBHB is not encoded solely in the characteristic strain of the GW signal, but also in its dephasing. Our results demonstrate that, even when the GW amplitude is strongly reduced over a certain frequency range due to the presence of high-density DM spikes, significant dephasing effects could still be present and observable by LISA. This, of course, assumes that the SMBHB evolution occurs within LISA’s sensitivity range and that the observation time is sufficiently long.

Furthermore, it is important to address that the formation mechanisms, evolution, and long-term stability of DM spikes around SMBHs are still subject to debate, especially regarding their inner density slopes and the effects of dynamical processes. Within the theoretical framework of dynamical friction, we do not explicitly consider the energy transfer within the spikes, which would be caused by the interaction of the DM particles with each other. In the case of self-interacting DM particles, this could lead to the smoothing of the density profile in the inner regions of the spike \cite{Elbert_2015}. Studying this effect in detail is beyond the scope of this work and is left for future research. Using a simplified estimate of the energy balance between the DM binding energy and the energy deposited by dynamical friction \cite{Kavanagh_2020}, we find that the existence of substantial CDM spikes around the SMBHs considered here appears very unlikely (see Appendix \ref{sec:energy_balance_CDM}). Nevertheless, this conclusion should be interpreted with caution, as more sophisticated numerical simulations would be required for a robust assessment \cite{Mukherjee_2023, Kavanagh_2025}.

An additional aspect to consider is that a relativistic treatment of spike formation around SMBHs results in an increased central spike density, as studied by Ref.~\cite{Sadeghian_2013}. However, since the innermost region contains only a small fraction of the total spike mass, the overall impact of this effect is limited. Moreover, for the reasons discussed in Sec.~\ref{sec:theory}, the numerical evolution of the SMBHBs is artificially terminated at $r = r_{\mathrm{ISCO},1} + r_{\mathrm{ISCO},2}$. Nevertheless, the higher central densities in the relativistic case lead to a slightly faster final inspiral of the SMBHBs. This, however, does not affect the upper detection limits for LISA, as they are model-independent.

In addition to LISA, there are further possibilities for SMBHB detection. In particular, microhertz frequencies would allow for the detection of more massive systems with $M \geq 10^8~\rm{M}_\odot$. Currently, the only bounds on GWs in this regime come from Doppler tracking of the \textit{Cassini} spacecraft \cite{Armstrong_2003}. While a dedicated laser interferometer at these frequencies would be beyond the reach of current or near-future technological capabilities \cite{Sesana2021}, there are several ideas for ways to detect GWs in this frequency regime. Some of the possible ideas include: using relative astrometry from the Nancy Grace Roman Space Telescope (NGRST) to detect GWs astrometrically \cite{Wang2021, Wang2022, Pardo2023}, measuring the modifications to asteroid accelerations induced by GWs \cite{Fedderke2022}, measuring the phase modulation that these GWs would induce in LISA data \cite{Bustamante-Rosell2022}, and monitoring binary systems for deviations from GWs \cite{Blas2022, Blas2022L}. Should any of these techniques make a detection of a binary system, the method outlined in this paper could be used to set bounds on the DM density near the SMBHs.

Finally, it should also be mentioned that an attenuation of the GW signal can also be observed for eccentric SMBHs without any surrounding matter. This was first investigated by Ref.~\cite{Enoki_2007} in the context of the stochastic GWB produced by an eccentric SMBHB population. It is already clear from this study that a high eccentricity is necessary for GW-driven systems so that the attenuation is visible within the PTA band. This was later confirmed by Ref.~\cite{Raidal_2024}, among others \cite{Chen_2017}. SMBHBs could exhibit such high eccentricities in stellar environments, since the gravitational slingshot ejection of stars or matter clumps through three-body interactions could lead to an eccentrification of the orbits \cite{Quinlan_1996, Sesana_2010}. The shape of the stochastic GWB generated by eccentric and slingshot ejection-driven SMBHBs was first explored by Ref.~\cite{Sesana_2013}. Refs.~\cite{Taylor_2017}, \cite{Kelley_2017} and \cite{Bonetti_2023} conducted further studies on this and found that it is difficult to disentangle the effects on the stochastic GWB caused by stellar hardening and eccentricity, as both lead to a similar attenuation of the characteristic strain at low frequencies. We explore this in more detail by using the data of the 15-year set by NANOGrav \cite{Agazie_2023_1} in Ref.~\cite{Chen_2024}. 

The main result of this study shows that high matter densities in SMBHBs significantly influences their orbital evolution and the GWs they emit, particularly within the frequency range accessible to LISA. Our model-independent approach offers a framework for estimating the matter density around SMBHs from GW observations or establishing upper detection limits for different detectors by using their sensitivity curves. Moreover, it indicates that LISA will be able to place constraints on the matter density near these black holes. Future observations with LISA or other detectors operating in the microhertz range will provide further insight into SMBHB environments and help to improve these density estimates.

\appendix
\section{Energy balance considerations of static CDM spikes \label{sec:energy_balance_CDM}}

To assess the stability of CDM spikes around the SMBHs considered in this work, we evaluate the balance between the gravitational binding energy of the spikes and the energy deposited into them by the inspiralling SMBHBs via dynamical friction. By closely following the approach of Ref.~\cite{Kavanagh_2020}, we outline the relevant expressions below. This represents a simplified analytical estimate that cannot replace numerical simulations, as already discussed in \ref{subsec:results_1} in detail.

\subsection{Potential binding energy of DM spikes \label{subsec:appendix_U_DM}}
For a spherically symmetric, static DM spike around a SMBH with mass $m_\mathrm{BH}$, as described in \ref{subsec:dm-spike}, the potential binding energy $U_\mathrm{DM}(r)$ of the spike particles within a radius $r$ is given by

\begin{align}
    U_\mathrm{DM}(r) = - \int_{r_\mathrm{min}}^r \frac{G[m_\mathrm{BH}+m_\mathrm{enc}(r')]}{r'} 4 \pi r'^2 \rho_\mathrm{DM}(r') dr'\;,
    \label{eq:U_DM(r)}
\end{align}

 \noindent where $\rho_\mathrm{DM}(r)$ is the spike density distribution (cf. Eq.~\ref{eq:rho_DM(r)}), and $m_\mathrm{enc}(r)$ denotes the enclosed DM mass within the radius $r$, which can be expressed as

 \begin{align}
     m_\mathrm{enc}(r) = \begin{cases}
                        0 & \text{, } r < r_{\text{min}} \\
                        m_\mathrm{DM}(r) - m_\mathrm{DM}(r_\mathrm{min}) & \text{, } r_h \geq r \geq r_{\text{min}} \\
                      \end{cases}\;,
    \label{eq:m_enc(r)}
 \end{align}

 \noindent with 

 \begin{align}
     m_\mathrm{DM}(r) = \frac{4 \pi \rho_\mathrm{sp} r^{\alpha}_\mathrm{sp}}{3-\alpha}\,r^{3-\alpha}\;.
     \label{eq:m_DM(r)}
 \end{align}

\noindent Carrying out the integration in Eq.~\ref{eq:U_DM(r)} yields

\begin{equation}
\begin{aligned}
     U_\mathrm{DM}(r) &= -\frac{G m_\mathrm{DM}(r)(3-\alpha)}{r} \\[0.2cm] &\times \left[\frac{m_\mathrm{BH}-m_\mathrm{DM}(r_\mathrm{min})}{2-\alpha}+\frac{m_\mathrm{DM}(r)}{5-2\alpha}\right] - U_\mathrm{min}\;,
     \label{eq:U_DM(2)-2}
\end{aligned}
\end{equation}

\noindent where the constant $U_\mathrm{min}$ is defined as

\begin{equation}
    \begin{aligned}
U_\mathrm{min} &= -\frac{G m_\mathrm{DM}(r_\mathrm{min})(3-\alpha)}{r_\mathrm{min}(2-\alpha)}\\[0.2cm] 
&\times \left[m_\mathrm{BH}-\frac{m_\mathrm{DM}(r_\mathrm{min})(3-\alpha)}{5-2\alpha}\right]\;.
    \label{eq:U_min}
\end{aligned}
\end{equation}

\noindent Here, we note again that $r_\mathrm{min}$ is equated with $r_\mathrm{ISCO}$.

Therefore, the total binding energy of the two DM spikes in a binary system with equal-mass SMBHs is $2\times U_\mathrm{DM}(r_h)$, assuming both spikes have the same slope parameter $\alpha$.

\subsection{Energy dissipation via dynamical friction \label{subsec:appendix_Delta_E_DF}}

\begin{figure}
    \centering
    \includegraphics[width=0.98\columnwidth]{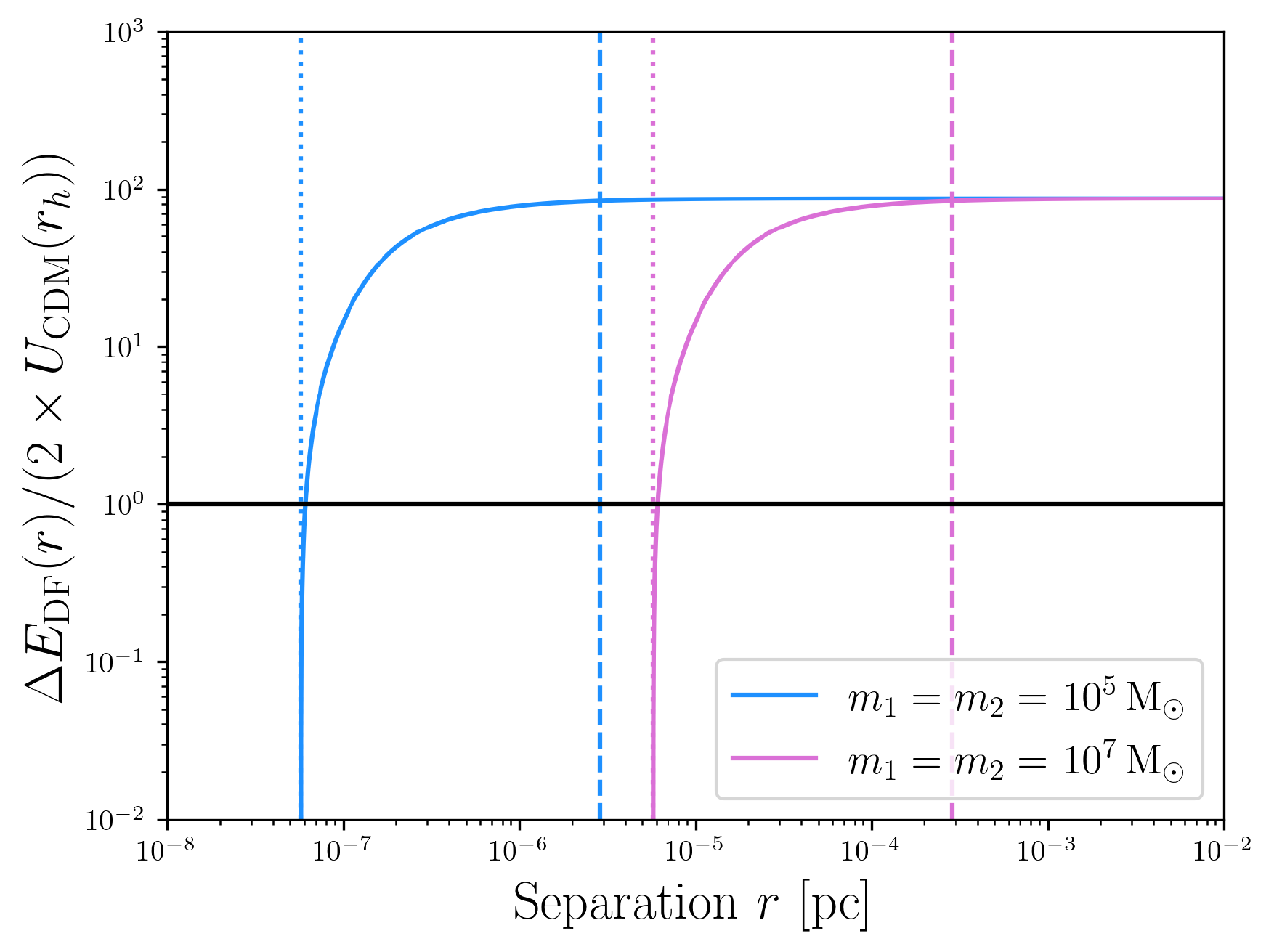}
    \caption{Ratio between the energy dissipated due to dynamical friction and the total binding energy of both CDM spikes in circular equal-mass SMBHBs with $m_1 = m_2 = 10^5\,\mathrm{M_\odot}, 10^7\,\mathrm{M_\odot}$ (blue, pink) as a function of separation $r$. The vertical dashed lines indicate the chosen initial separation $r_0 = 50\times(r_\mathrm{ISCO,1}+r_\mathrm{ISCO,2})$, while the vertical dotted lines represent $r_\mathrm{fin} = r_\mathrm{ISCO,1}+r_\mathrm{ISCO,2}$ for each SMBHB.\newline $\{\sigma,\, \ln(\Lambda)\} = \{200\,\text{km/s},\,10\}$}
    \label{fig:Delta_E_DF_vs_U_CDM}
\end{figure}

The total energy $\Delta E_\mathrm{DF}(r)$ dissipated due to dynamical friction as a circular SMBHB inspirals from an arbitrary separation $r$ down to $r_\mathrm{fin} = r_\mathrm{ISCO,1}+r_\mathrm{ISCO,2}$ is given by
\begin{align}
    \Delta E_\mathrm{DF}(r) = \int_r^{r_\mathrm{fin}} \frac{dE_\mathrm{DF}}{dt}\,\left(\frac{dr'}{dt}\right)^{-1}\,dr'\;,
    \label{eq:DeltaE_DF(r)}
\end{align}

\noindent with $dr/dt$ provided by Eq.~\ref{eq:da_dt}. For a SMBHB consisting of equal-mass black holes, $m_1 = m_2 = m_\mathrm{BH} = m$, and considering identical spike slopes, Eq.~\ref{eq:dE_DF_} can be rewritten as
\begin{align}
    \frac{dE_\mathrm{DF}}{dt} = - 4 \pi G^2 \ln(\Lambda) \rho_\mathrm{DM}(r) \frac{m^2}{v}
    \label{eq:dE_DF_equal_masses}
\end{align}

Inserting this into Eq.~\ref{eq:DeltaE_DF(r)}, the integral can be expressed using hypergeometric functions

\begin{equation}
   \begin{aligned}
    \Delta E_\mathrm{DF}(r) & = \biggl[\frac{Gm^2}{2r'} \\[0.2cm]
    &\times{}_2F_1\left(1,\frac{2}{11-2\alpha};\frac{13-2\alpha}{11-2\alpha};-c_r\, r'^{-11/2+\alpha}\right)\biggl]_{r_\mathrm{fin}}^{r}
    \label{eq:DeltaE_DF(r)_eval}
    \end{aligned} 
\end{equation}

\noindent with 

\begin{align}
    c_r = \frac{2^{9/2}G^{5/2}m^{7/2}}{5\pi c^5 \rho_\mathrm{sp} r_\mathrm{sp}^{\alpha} \ln(\Lambda)}
\end{align}

\subsection{Ratio between energy dissipated and binding energy}
By evaluating Eqs.~\ref{eq:U_DM(2)-2} and \ref{eq:DeltaE_DF(r)_eval} for CDM spikes with $\alpha = 7/3$, we find that the ratio $\Delta E_\mathrm{DF}(r)/(2\times U_\mathrm{CDM}(r_h))$ already exceeds unity at $r\approx r_\mathrm{fin}$ for both considered equal-mass SMBHBs, as shown in Fig.~\ref{fig:Delta_E_DF_vs_U_CDM}. At the initial separation, this ratio almost reaches its maximum of about 87. Consequently, the CDM spikes are significantly disrupted and are rather unlikely to survive around the SMBHs studied in this work. We emphasize again that this conclusion relies on several simplifying assumptions. For more realistic SMBHB environments, or when using more advanced treatments such as N-body simulations, the outcome could be very different, as already indicated by Refs.~\cite{Mukherjee_2023, Kavanagh_2025}.

In contrast, Ref.~\cite{Alonso-Alvarez_2024} suggests that the collisional nature of SIDM may enable the particles to efficiently repopulate SIDM spikes, thereby counteracting the destruction and potentially sustaining the spikes over extended timescales. Studying this effect in detail lies beyond the scope of this work. Nevertheless, we provide a simple timescale comparison to support this statement in the following section.

\section{Stability of SIDM spikes: Timescale comparison \label{sec:timescale_estimate_SIDM}}

To assess the potential stability of SIDM spikes surrounding SMBHs, we compare two characteristic timescales:

\begin{itemize}
    \item The \textit{dynamical friction timescale} $t_\mathrm{DF}$, defined as the time it takes a SMBHB to shrink its orbital separation from some initial value $r_0$ down to a smaller value $r \geq r_\mathrm{ISCO,1}+r_\mathrm{ISCO,2}$, solely due to energy loss via dynamical friction with SIDM spikes.
    \item The \textit{replenishment timescale} $t_\mathrm{replenish}$, which quantifies the rate at which SIDM particles repopulate the spike at $r$ due to their self-interacting nature.
\end{itemize}

For circular equal-mass SMBHBs with $m_1 = m_2 = m_\mathrm{BH}$, the inspiral timescale from $r_0 = 50\times (r_\mathrm{ISCO,1}+r_\mathrm{ISCO,2})$ to $r$ is given by
\begin{equation}
   \begin{aligned}
    t_\mathrm{DF} &= \int_{r_0}^{r} \left(\frac{dr'}{dt}\biggl|_\mathrm{DF}\right)^{-1}\,dr'\\[0.3cm]
    &= \frac{m_\mathrm{BH}^{1/2} \left(r_0^{\alpha-3/2}-r^{\alpha-3/2}\right)}{2^{5/2}\pi G^{1/2} \ln(\Lambda) \rho_\mathrm{sp}r^\alpha_\mathrm{sp} (\alpha-3/2)}\;,
    \label{eq:t_DF}
    \end{aligned} 
\end{equation}
\noindent where $dr/dt|_\mathrm{DF} = dE_\mathrm{DF}/dt\,\left(dE_\mathrm{orb}/dr\right)^{-1}$ is used (see Eq.~\ref{eq:da_dt}). 

The energy released by the SMBHB through dynamical friction is expected to be absorbed by the SIDM spikes, thereby reducing their potential binding energies. Therefore, it is reasonable to assume that the spikes' destruction occurs on the same timescale $t_\mathrm{DF}$. Consequently, if $t_\mathrm{replenish} \ll t_\mathrm{DF}$, the SIDM particles replenish the spike sufficiently fast before significant energy deposition disrupts its structure. In this case, the SIDM spike can be considered dynamically stable during the inspiral of the binary.

The replenishment timescale for a SIDM spike can be estimated by \cite{BinneyTremaine_2008, Almeida_2021}
\begin{align}
    t_\mathrm{replenish} = \left(\frac{\sigma_p}{m_p}\, \rho_\mathrm{DM}(r) \, v_p(r)\right)^{-1}\;,
\label{eq:t_replenish}
\end{align}
\noindent where $v_p(r) = \sqrt{G[m_\mathrm{BH}+m_\mathrm{enc}(r)]/r}$ is the virial velocity of SIDM particles at $r$ and $\sigma_p/m_p$ denotes the self-interaction cross section per unit mass. The enclosed DM mass $m_\mathrm{enc}(r)$ is defined in Eq.~\ref{eq:m_enc(r)}.

By setting $\sigma_p/m_p = 1\,\mathrm{cm^2/g}$, $\alpha = 7/4$, $\sigma = 200\,\mathrm{km/s}$ and $\ln(\Lambda) = 10$, we find $t_\mathrm{DF} \gtrsim 10^8\,t_\mathrm{replenish}$ for $m_\mathrm{BH} = 10^5\,\mathrm{M_\odot}$ and $t_\mathrm{DF} \gtrsim 10^6\,t_\mathrm{replenish}$ for $m_\mathrm{BH} = 10^7\,\mathrm{M_\odot}$, respectively. For other values of $\alpha \neq 3/2$ (see \ref{subsec:dm-spike}), the corresponding ratios between $t_\mathrm{DF}$ and $t_\mathrm{replenish}$ remain at a similar level.

\bibliographystyle{apsrev4-1}

\end{document}